\documentclass[pra,aps,showpacs,twocolumn,floatfix]{revtex4}
\usepackage{graphicx}
\usepackage{array}
\usepackage{color}
\usepackage{amsmath}
\usepackage{amsxtra}
\usepackage{amstext}
\usepackage{amssymb}
\usepackage{latexsym}
\usepackage{dsfont}

\begin{document}

\title{Photon-phonon parametric oscillation induced by the quadratic coupling in an optomechanical resonator}
\author{Lin Zhang}
\email{zhanglincn@snnu.edu.cn}
\author{Fengzhou Ji}
\author{Xu Zhang}
\affiliation{School of Physics and Information Technology, Shaanxi Normal University,
Xi'an 710119, China}
\author{Weiping Zhang}
\affiliation{Quantum Institute for Light and Atoms, Department of Physics,
East China Normal University, No.500, Shanghai 200241, P. R. China}


\begin{abstract}
A direct photon-phonon parametric effect of the quadratic coupling on the mean-field dynamics
of an optomechanical resonator in the large-scale-movement regime is found and investigated. Under a
weak pumping power, the mechanical resonator damps to steady state with a
nonlinear static response sensitively modified by the quadratic coupling. When the driving power
increases beyond the static energy balance, the steady states lose their stabilities
via Hopf bifurcations and the resonator produces stable self-sustained oscillation
(limit-circle behavior) of discrete energies with step-like amplitudes due to the parametric
effect of the quadratic coupling, which can be understood roughly by the power balance between gain and loss
on the resonator. A further increase of the pumping power can induce
chaotic dynamic of the resonator via a typical routine of period-doubling bifurcation
but which can be stabilized by the parametric effect through an inversion bifurcation process back
to limit-circle states. The bifurcation-to-inverse-bifurcation transitions
are numerically verified by the maximal Lyapunov exponents of the dynamics and
which indicate an efficient way to suppress the chaotic behavior of the optomechanical
resonator by the quadratic coupling. Furthermore, the parametric effect of the quadratic
coupling on the dynamic transitions of an optomechanical resonator can be conveniently
detected or traced by the output power spectrum of the cavity field.
\end{abstract}

\maketitle

\section{Introduction}

In recent years, continuing interests on cavity optomechanical systems stimulate the study
of light driven nanomechanical resonators in order to develop high-efficiency nano-motors,
ultra-sensitive mass/force sensors or high-speed, low-energy consuming signal processors \cite{Marqu,Clerk}. In
a red-detuned weak-driving regime, the quantum behaviors of nanomechanical resonators is the main
topic of research, such as mechanical cooling, quantum state controlling, state engineering
and state transferring, etc \cite{Poot,Meystre}. However, under a strong or in a blue-detuned pumping
field, a large-scale movement up to micrometers can be induced to evoke strong nonlinear
properties of the nanomechanical resonators that invalidate the harmonic oscillator model.
One of the important nonlinear effects emerged in a large-scale movement is the quadratic coupling, a second
order photon-phonon coupling in the optomechanical systems \cite{Thompson,Chang,Sankey,Peano,Purdy,Gupta}.
The nonlinear behavior with a large-amplitude of motion is dominated by the mean-field dynamics
of the resonator and it is critical to explore new features of practical applications. One important
nonlinear behavior found in the optomechanical systems is that the system can support stable
self-sustained oscillations (SSOs) \cite{Metzger,Carmon2,Okamoto,Zaitsev} and a high-efficiency
broadband high-order harmonic oscillation can be generated in this case \cite{Lin,Kippenberg}. An easily controlled SSO
working at different power levels is the physical foundation for many practical applications such as developing force
sensors or information processors based on nonlinear photon-phonon interactions \cite{Jenkins}. Therefore, in this work,
we will explore a photon-phonon parametric oscillation of the optomechanical resonator by considering its quadratic
coupling in order to facilitate an adaptive application of this system in the future.

As the typical displacement of a nano-resonator is extremely small, the conventional optomechanical model
considers only the linear photon-phonon coupling, whose magnitude is linearly  proportional to the
displacement of the mechanical resonator, and the quadratic coupling is very weak then. However,
for a mechanical resonator in an extremely-large-amplitude regime \cite{Long}, the weak quadratic coupling
starts to play non-neglectable roles because its magnitude is proportional to the displacement squared.
On the other hand, the chaotic behaviors of an optomechanical
resonator in red-detuned pumping fields have been extensively verified in many
optomechanical systems \cite{Bakemeier,Ying,Girvin,Carmon,Larson}.
Recently, Bakemeier et al. \cite{Bakemeier} studied the typical routes to
chaos of a linear coupling optomechanical resonator in the weak pumping fields.
They demonstrated a period-doubling bifurcation leading to chaos
within a small displacement region which greatly limits the physical applications of
optomechanical resonators due to the dynamical instabilities. Therefore, for a chaotic resonator, a very weak
quadratic coupling will play a critical role in the nonlinear dynamics \cite{Karabalin}.
As have been shown in many optomechanical systems \cite{Thompson,Chang,Sankey,Peano,Purdy,Gupta},
the quadratic phonon-photon couplings bring manifest nonlinear properties even in a weak red-detuned
pumping case and can lead to interesting parametric dynamics in a variety of optomechanical systems
\cite{Girvin0,Heidmann,Aldridge,Katz,Venstra}.

Based on the above discussions, we will closely consider the new dynamical features
of a parametric optomechanical resonator induced by the quadratic couplings in a strong red-detuned pumping field.
We find a direct parametric effect of the quadratic coupling on the dynamical properties
of the optomechanical resonator which is directly controlled by the intensity of the cavity field.
The dynamic behavior modified by the parametric effect provides a solid way to stabilize the mechanical
response of a nano-mechanical resonator in a large-amplitude regime, especially within the chaotic region, and the nonlinear properties
induced by the quadratic coupling can provide an easily controlled SSO with an enhanced
sensitive response to an external driving field in the critical parametric region. The combination of linear
and quadratic couplings provides a tunable nonlinearity of the nano-resonators which can be exploited to
develop the reliable sensitive on-chip signal processors or the tunable mechanical sensors \cite{Aldridge}.

\section{The parametric equation of the cross coupling resonator}

The Hamiltonian of an optomechanical oscillator with both linear and quadratic photon-phonon
couplings can be written as \cite{Marqu,Lin}
\begin{equation}
\hat{H}=\hat{H}_{\mathrm{M}}+\hat{H}_{\mathrm{cav}}+\hat{H}_{\mathrm{pump}}+%
\hat{H}_{\kappa }+\hat{H}_{\gamma},
\label{H0}
\end{equation}%
where
\begin{eqnarray}
\hat{H}_{\mathrm{M}} &=&\frac{\hat{p}^{2}}{2M}+\frac{1}{2}M\omega _{M}^{2}%
\hat{x}^{2},  \notag \\
\hat{H}_{\mathrm{cav}} &=&-\hbar \delta_c \hat{a}^{\dag }\hat{a}+\hbar G_1 \hat{a}%
^{\dag }\hat{a}\hat{x}+\frac{1}{2}\hbar G_2 \hat{a}^{\dag }\hat{a}\hat{x}%
^{2} \notag \\
\hat{H}_{\mathrm{pump}} &=&i\hbar \left[ \eta \left( t\right) \hat{a}%
^{\dagger }-\eta ^{\ast }\left( t\right) \hat{a}\right] . \notag
\end{eqnarray}
In the above model, the mechanical resonator $\hat{H}_{\mathrm{M}}$ with
a free frequency of $\omega_M$ couples with a cavity mode $\hat{H}_{%
\mathrm{cav}}$ of frequency $\omega_c$ through a linear pressure coupling rate
of $G_1$ and a dispersive quadratic coupling rate of $G_2$ \cite{Thompson}.
The high-order quadratic coupling can be effectively derived from not only the nonlinear
deformation of the cavity field but also the nonlinear
properties of the mechanical resonators \cite{Peano,Favero,Zaitsev}.
$\hat{H}_{\mathrm{pump}}$ describes the cavity mode pumping
by a classical field with a frequency of $\omega_p$ and an amplitude of $\eta $ related
to its pumping power $P$ by
\begin{equation}
\label{power}
 P=|\eta |^{2}\hbar \omega_{p}/\kappa.
\end{equation}
$\hat{H}_{\kappa }$ and $\hat{H}_{\gamma}$ are the baths coupled to the cavity mode and
the mechanical resonator and induce energy dissipations with the damping rates of $\kappa$
and $\gamma_M$, respectively. The pump-cavity detuning is defined by $\delta _{c}=\omega_{p}-\omega _{c}$,
and the coupling rates of $G_1$, $G_2$ are given by $G_1=\omega_c^{\prime}(x)$, $G_2 =\omega_c^{\prime\prime}(x)$
for a linear cavity, where the prime means the spatial derivatives \cite{Marqu}.

Based on the cross-coupling Hamiltonian of Eq.(\ref{H0}), the Heisenberg equation of motion
for the mechanical resonator leads to a parametric oscillator described by
\begin{equation}
\frac{d^{2}\hat{x}}{dt^{2}}+\frac{\gamma _{M}}{2}\frac{d\hat{x}}{dt}+\omega
_{M}^{2}\left( 1+\frac{\hbar G_2 }{M\omega _{M}^{2}}\hat{a}^{\dagger }%
\hat{a}\right) \hat{x}=-\frac{\hbar G_1}{M}\hat{a}^{\dagger }\hat{a}+\frac{%
\hat{\xi}\left( t\right) }{M},
\label{dyn1}
\end{equation}%
where $\hat{\xi}(t)$ accounts for the Langevin noise of the mechanical
resonator. Clearly, Eq.(\ref{dyn1}) indicates that the quadratic coupling
introduces a direct parametric modification of the frequency of the
resonator which consistently controlled by the
cavity field intensity of $\hat{a}^{\dag}\hat{a}$. For a conventional system with $G_2=0$,
Eq.(\ref{dyn1}) reduces to
\begin{equation*}
\frac{d^{2}\hat{x}}{dt^{2}}+\frac{\gamma _{M}}{2}\frac{d\hat{x}}{dt}+\omega
_{M}^{2}\hat{x}=-\frac{\hbar G_1}{M}\hat{a}^{\dagger }\hat{a}+\frac{%
\hat{\xi}\left( t\right) }{M},
\end{equation*}%
which only includes a pressure force induced by the cavity field through the linear coupling
of $G_1$. Therefore, the quadratic coupling $G_2$ in Eq.(\ref{dyn1}) introduces further a parametric effect
on the frequency of the resonator directly related to the intensity of the cavity field.
The intensity of the cavity mode is then determined by
\begin{equation*}
\frac{d}{dt}\hat{a}^{\dagger }\hat{a}=\eta \hat{a}^{\dagger }+\eta ^{\ast }%
\hat{a}-\kappa \hat{a}^{\dag }\hat{a}+\sqrt{\kappa }\left( \hat{a}^{\dag }%
\hat{a}_{in}+\hat{a}_{in}^{\dag }\hat{a}\right),
\end{equation*}
where $\hat{a}_{in}$ describes the input noise of the light mode and the coupling rate
$\sqrt{\kappa}$ is due to the cavity input-output relation \cite{Walls}.
Therefore, the optomechanical
resonator with both linear and quadratic couplings
results in a field-driven parametric oscillator \cite{Rugar,Turner,Junho} with
field-dependent frequency and driving force.
Although the quadratic coupling is weak, its parametric
effect on the mechanical oscillation will definitely bring new dynamic features which
would become manifest when the amplitude of the motion becomes large and the intensity
of the cavity field grows stronger. In this case,
the mean-field motion will dominate the dynamics by masking all the quantum correlations of the system.
Therefore the behavior of the resonator can be investigated
by the mean values of the quantum operators in the sense of $c$ numbers, such as $\hat{x}\rightarrow \langle\hat{x}%
\rangle=x_{c},\hat{p}\rightarrow\langle\hat{p}\rangle= p_{c},\hat{a}%
\rightarrow\langle\hat{a}\rangle=a_{c}$ without considering any quantum fluctuations
and correlations \cite{Ludwig}. Therefore the dynamics of the system is governed by the following
system of equations:%
\begin{eqnarray}
\frac{dx\left( \tau \right) }{d\tau } &=&p\left( \tau \right) ,  \notag \\
\frac{dp\left( \tau \right) }{d\tau } &=&-\frac{\gamma _{M}}{2}p-\Omega_{c}
^{2}\left( \tau \right) x\left( \tau \right) -2g_{1}I\left( \tau
\right)   \notag \\
\frac{dX\left( \tau \right) }{d\tau } &=&-\left[ \delta _{c}-g_{1}
x\left( \tau \right) -g_{2}x^{2}\left( \tau \right) \right]
Y\left( \tau \right)   \label{dyn} \\
&&-\frac{\kappa }{2}X\left( \tau \right) +\eta \left( \tau \right) ,  \notag
\\
\frac{dY\left( \tau \right) }{d\tau } &=&\left[ \delta _{c}-g
_{1}x\left( \tau \right) -g_{2}x^{2}\left( \tau \right) \right]
X\left( \tau \right) -\frac{\kappa }{2}Y\left( \tau \right) ,  \notag
\end{eqnarray}%
where $x(\tau) =x_{c}/x_{\mathrm{zpt}}$ and
$p\left( \tau \right) =p_{c}/p_{\mathrm{zpf}}$ are the position and
the momentum of the mechanical resonator scaled by $x_{\mathrm{zpf}}=\sqrt{\hbar /2M\omega _{M}}$
and $p_{\mathrm{zpf}}=\sqrt{\hbar M\omega _{M}/2}$, respectively.
The two field quadratures $X\left( \tau \right) $ and $Y\left( \tau \right) $ are
defined by $X\left( \tau \right) =\left( a_{c}+a_{c}^{\ast
}\right) /2$ and $Y\left( \tau \right) =-i\left( a_{c}-a_{c}^{\ast }\right) /2$
with the field intensity being $I\left( \tau \right) =X^{2}\left( %
\tau \right) +Y^{2}\left( \tau \right)$ and the field-modulated frequency of the
mechanical resonator is defined by $\Omega_{c}
\left( \tau \right) =\sqrt{1+4g_{2}I\left( \tau \right) }$.
Particularly, the scaled linear and quadratic coupling rates are%
\begin{equation}
g_{1}=G_1 x_{\mathrm{zpf}}/\omega _{M}, g_{2}=\frac{1}{2}G_2 x_{%
\mathrm{zpf}}^{2}/\omega _{M},  \label{parameter}
\end{equation}%
respectively, and all the other parameters $\delta_c, \gamma_M, \kappa$ are
scaled by $\omega_M$ for convenience with a scaled
time of $\tau =\omega _{M}t$.
In the above equations, the mean-time influence of the equilibrium
baths $\hat{H}_\kappa$ and $\hat{H}_\gamma$ are supposed to be zero for $\left\langle
a_{in}\left( \tau \right) \right\rangle =0$ and $\left\langle \xi \left( \tau \right) \right\rangle =0$
in a classical time limit. The amplitude of the classical pumping field is supposed to be a
real function of $\eta \left( \tau\right) $ and also scaled by $\omega _{M}$.

Then the classical version of Eq.(\ref{dyn1}) becomes%
\begin{equation}
\ddot{x}\left( \tau \right) +\frac{\gamma _{M}}{2}\dot{x}\left( \tau \right)
+\Omega_{c} ^{2}\left( \tau \right) x\left( \tau \right) =-2g_{1}I\left(
\tau \right) .  \label{cdeq}
\end{equation}%
Eq.(\ref{cdeq}) presents a typical parametric equation which enables a parametric
driving of the resonator by the cavity field through the quadratic coupling, and, reversely,
the motion of the mechanical resonator also provides a parametric modification on the cavity field's
frequency. The quadratic coupling here converts a conventional resonator into a direct parametric
oscillator and can bring new dynamic features of the system such as parametric amplification,
resonance and stabilization. Uniquely, in the present case, the parametric effects are consistently
controlled by the cavity field which also simultaneously produces a light pressure force on the resonator.
Specifically, if the cavity field adaptively locks into a periodic motion, Eq.(\ref{cdeq}) will be a
periodically driven parametric oscillator and the Floquet's theorem or the properties of
Mathieu equation can reveal a rich dynamical behaviors of the optomechanical resonator.
As Eq.(\ref{dyn}) generally describes a parametric oscillator with a field-controlled frequency
through the quadratic coupling, a slight modification on the quadratic coupling rate can influence
the parametric process of the resonator by a consistent cavity field.
In the following analysis we will adopt the parameters that are closely relevant to the experimental results
and have been analyzed in detail in Ref.\cite{Sankey,Lin}.

\section{Parametric effect on static responses}

In a weak pumping field, the optomechanical resonator finally
settles down to a steady state due to the dominate role of energy damping.
The static response of the resonator is then determined by the steady states of \cite{Lin2}
\begin{eqnarray}
x_{s} &=&-\frac{2g_{1}I_{s}}{1+4g_{2}I_{s}},
\label{stable1} \\
I_{s} &=&\frac{\eta^{2}}{\left( \kappa /2\right) ^{2}+\left( \delta
_{c}-g_{1}x_{s}-g_{2}x_{s}^{2}\right) ^{2}},%
\label{stable2}
\end{eqnarray}%
where $I_{s}\equiv X_{s}^{2}+Y_{s}^{2}$ is the steady intensity of the cavity
field and $\eta^{2}$ is the scaled input pumping power (see Eq.(\ref{power})).
Then the static displacement responses of the resonator are determined by a quintic
equation of
\begin{eqnarray}
\left( \delta
_{c}-g_{1}x_{s}-g_{2}
x_{s}^{2}\right) ^{2}x_{s}&+&\left( \kappa/2\right)^{2} x_{s} \notag \\
&+&2\eta^{2}\left( g_{1}
+2g_{2}x_{s}\right) =0.
\label{xs}
\end{eqnarray}%
Eq.(\ref{xs}) reveals a clear nonlinear response of the resonator
to the light field \cite{Kozinsky,Dorsel}
and its analytical solutions are impossible to obtain.
As the quadratic coupling is very weak for $g_2 \ll 1$, Eq.(\ref{xs}) can
be simplified to a quartic equation
\begin{equation}
a_{4}x_{s}^{4}+a_{3}x_{s}^{3}+a_{2}x_{s}^{2}+a_{1}x_{s}+a_{0}=0,
\label{quartic}%
\end{equation}%
where the coefficients are $a_{4} =2g_{1}g_{2}$,
$a_{3} =g_{1}^{2}-2 g_{2}\delta _{c}$,
$a_{2} =-2 g_{1}\delta _{c}$,
$a_{1} =\delta _{c}^{2}+( \kappa /2) ^{2}+4g_{2}\eta
^{2}$, $a_{0} =2g_{1}\eta^{2}$.
\begin{figure*}[htp]
\includegraphics[width=0.32\textwidth]{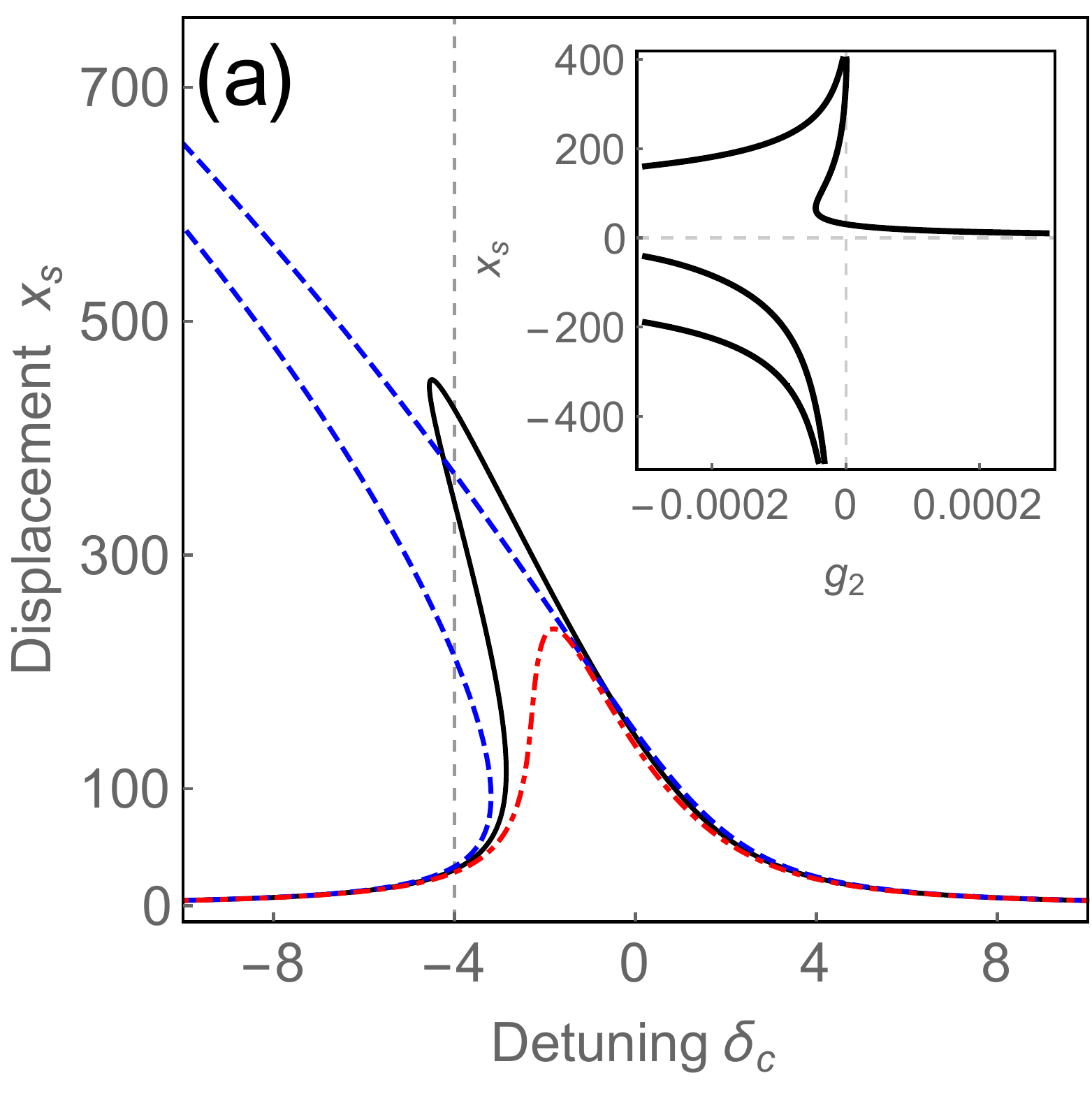}%
\includegraphics[width=0.33\textwidth]{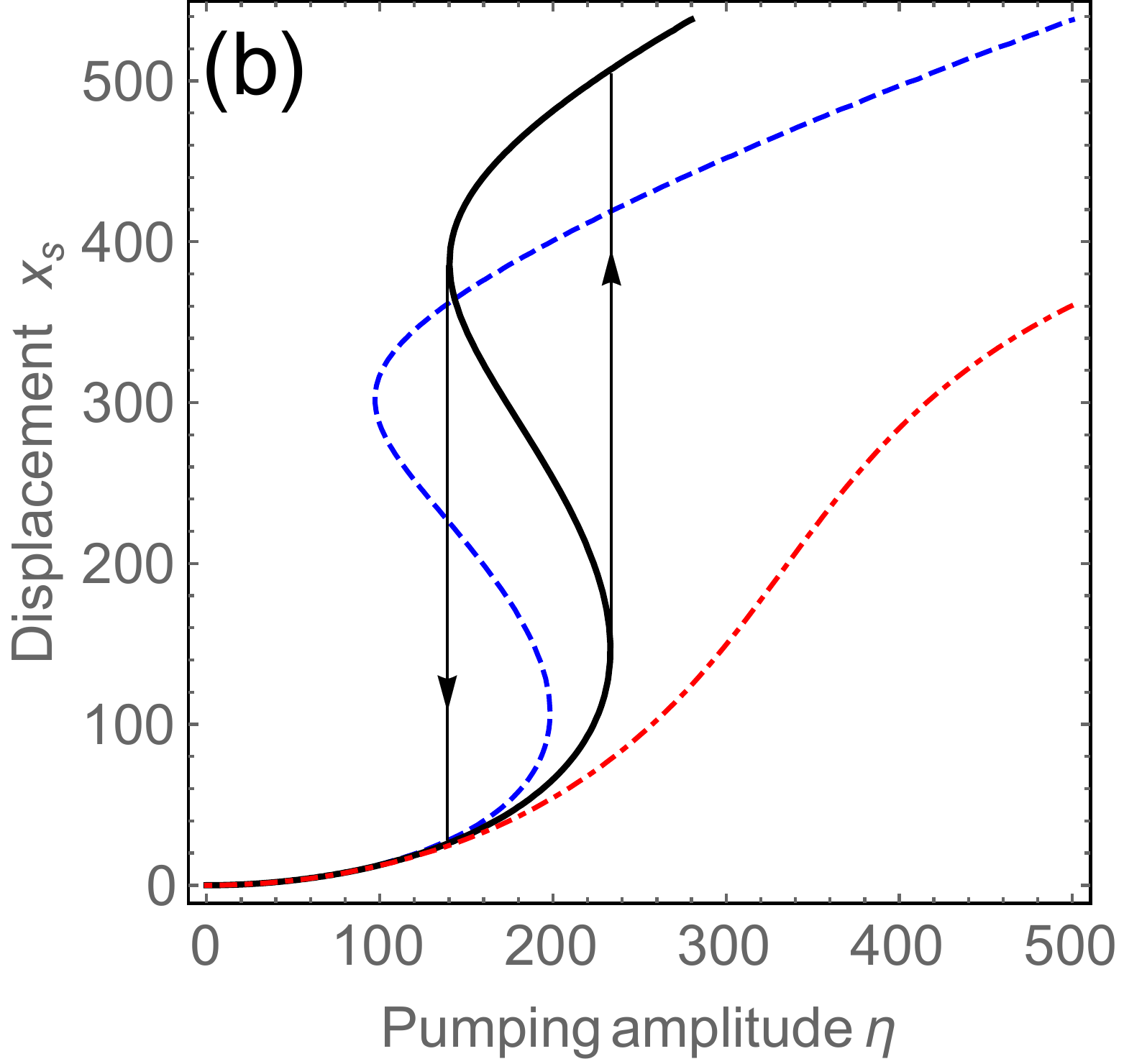}%
\includegraphics[width=0.33\textwidth]{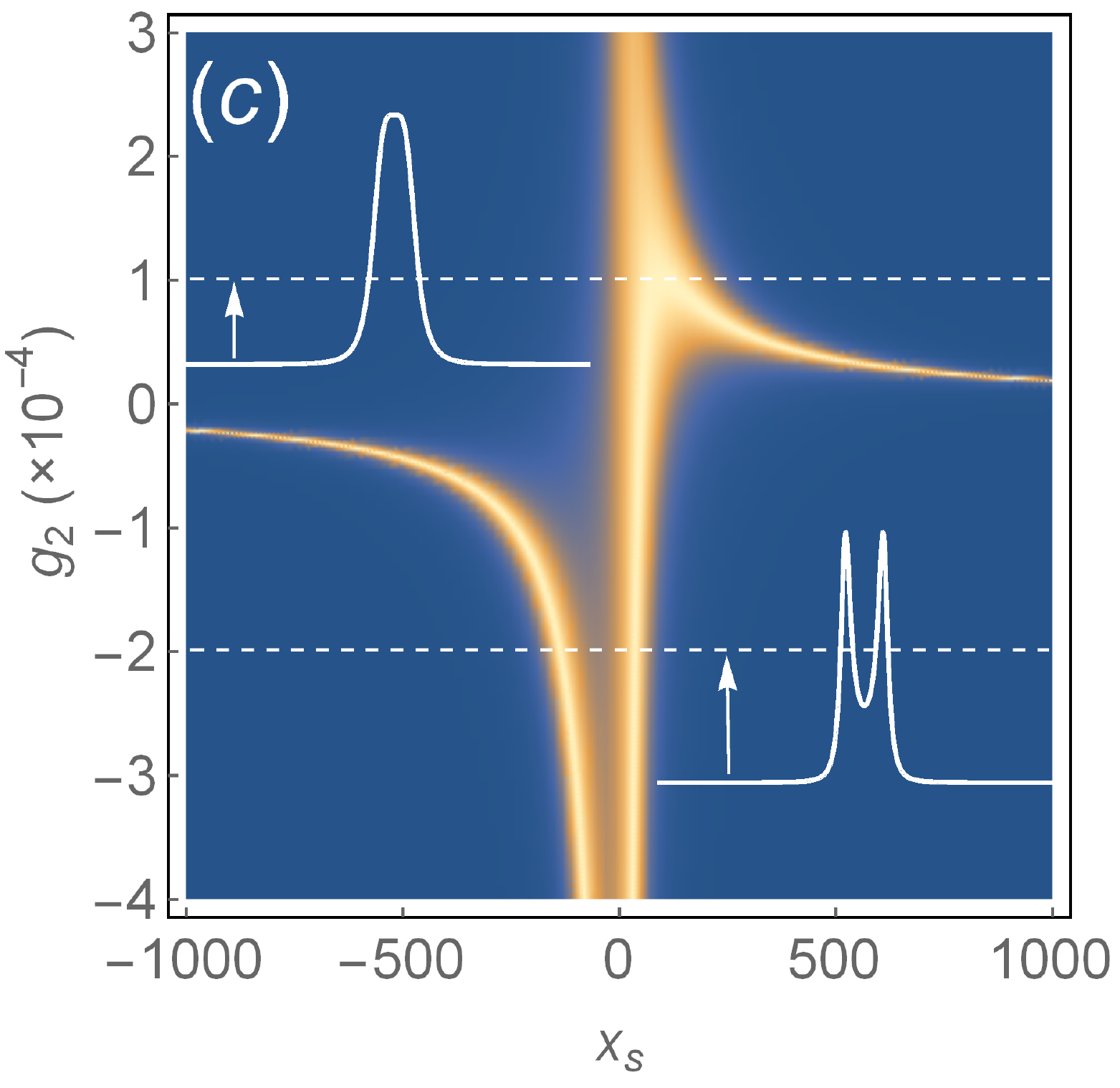}%
\caption{(a) Quadratic modification on static displacement $x_{s}$
versus pump detuning $\delta _{c}$ with a fixed pumping power (amplitude $\eta=150$).
Inset: $x_{s}$ changes with the quadratic coupling under
$\delta_c=-4$; (b) The bistability responses of $x_s$
to pumping amplitude $\eta$ with a fixed detuning of $\delta_c=-4$.
In both (a) and (b), the linear coupling $g_{1}=-0.01$ and
the quadratic couplings are $g_{2}=-10^{-5}$ (blue dashed lines),
$g_{2}=0$ (black solid lines), $g_{2}=+10^{-5}$ (red dot-dashed lines), respectively;
(c) The intensity response of the intracavity field, $I_s(g_2,x_s)$, to the position of
the resonator $x_s$ and the quadratic coupling $g_2$, where $\delta_c=-1$ and $g_1=-0.02$.
For all the above cases, the damping rate of the cavity field is $\kappa =2$.}
\label{figure1}
\end{figure*}%
Although the solutions of Eq.(\ref{quartic}) are now available, an exact
analysis on the static behaviors of $x_s$ is still cumbersome.  However, if both coupling rates
$|g_{1,2}|<10^{-2}$, the sign of discriminant of Eq.(\ref{quartic}) is determined by
$- g_2\delta_c$, which implies that the number of equilibria
of $x_s$ is mainly controlled by the detuning $\delta_c$
and the quadratic coupling $g_2$ \cite{Lin2}. An exact numerical analysis on Eq.(\ref{xs}) is shown in
Fig.\ref{figure1}, which indicates that the static response of the resonator is very
sensitive to the sign of quadratic coupling under a strong pumping \cite{Zaitsev}.
Fig.\ref{figure1} reveals that the negative quadratic couplings enhance the
nonlinear property of the mechanical resonator and the positive ones
can remove the bistability response of the resonator to the pumping field.
According to Eq.(\ref{stable1}), the static displacement of the resonator
approximately increases with the intensity of the cavity field due to
the linear coupling $g_1$, i.e. $\propto-g_1 I_s$,
and the extreme shift is determined only by the ratio of two coupling rates $-g_1/2g_2$.
Although in most cases the quadratic coupling $g_{2}$ is much smaller
than the linear coupling $g_{1}$, it brings dramatic effects
on the mechanical response if it meets a resonant case of $g_{2}\sim-1/4I_s$
demonstrated by the inset of Fig.\ref{figure1}(a). Eq.(\ref{stable2}), which is
shown in Fig.\ref{figure1}(c), reveals a double resonant peak of the intracavity intensity
to the resonator's position and the splitting of a normal Lorentzian lineshape of the cavity field
is $\sqrt{g^{2}_1+4\delta_c g_2}/g_2$, which is highly sensitive to the quadratic coupling.
Therefore, the splitting peak provides a method to detect the quadratic coupling rate of an
optomechanical resonator in the large-scale displacement regime.

\begin{figure}[htp]
\includegraphics[width=0.45\textwidth]{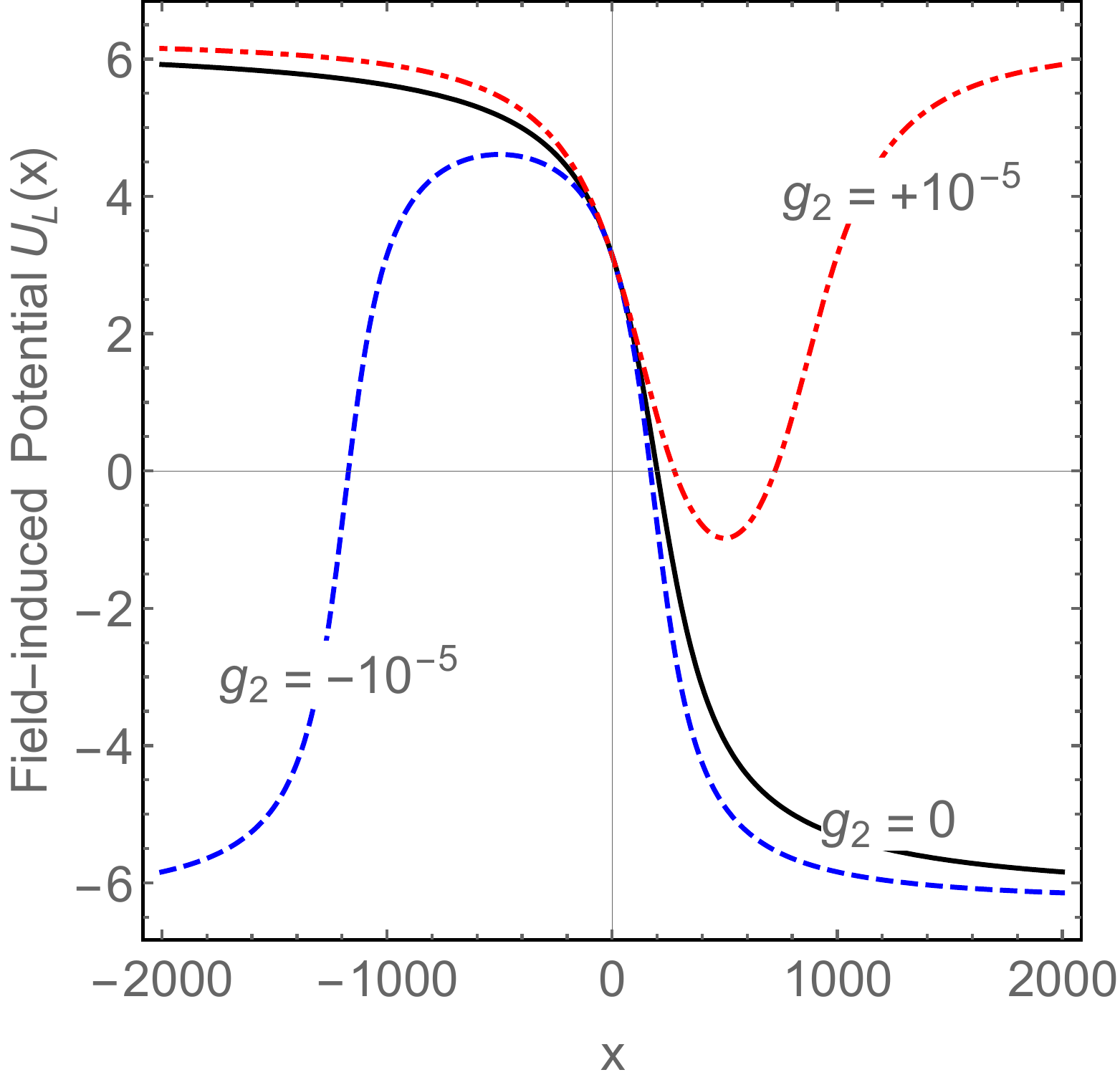}%
\caption{The light-induce potential $U_L(x)$ without (black solid line) and
with negative (blue dashed line) and positive (red dot-dashed line) quadratic couplings.
The other parameters are $\eta=2, \delta_c=-2, \kappa =4, g_{1}=-0.01$.}
\label{figure2}
\end{figure}

The nonlinear static responses modified by the quadratic coupling can be understood by the adiabatic
potential generated by the cavity mode when the field modulation is much faster
than the motion of the mechanical resonator. Then the total adiabatic potential felt by the
resonator is \cite{Lin}
\begin{equation}
U\left( x,\tau \right) =\frac{1}{2}x^{2}-\frac{4\eta ^{2}\left( \tau
\right) }{\kappa }\arctan \left[ \frac{\Delta \left( x \right) }{\kappa
/2}\right] ,  \label{U}
\end{equation}%
where the effective position-dependent detuning
\begin{equation}
\Delta \left( x \right) =\delta _{c}-g_{1}x-g
_{2}x^{2}.
\end{equation}
The light-dressed adiabatic potential,
\begin{equation}
U_L(x)\equiv-\frac{4\eta ^{2}}{\kappa }\arctan \left[ \frac{\Delta(x) }{\kappa
/2}\right],%
\label{UL}
\end{equation}
is an arctangent function which will exhibit distinct properties due to the quadratic
coupling as shown in Fig.\ref{figure2} (where $\eta=2$ is only for a convenient display).
The light-induced potential $U_L(x)$ will heavily dress the harmonic potential especially under
a large pumping power of $\eta^2$. The adiabatic potential shown in Fig.\ref{figure2}
indicates that a negative quadratic coupling can enhance the nonlinear properties
by introducing multiple stabilities and the positive one can suppress the nonlinear properties
by enhancing the trapping force of the harmonic potential.

However, all the above static analysis and the responsive method to detect the quadratic coupling in Fig.\ref{figure1}
are based on the stability of the steady states of Eq.(\ref{stable1}) and Eq.(\ref{stable2}), which
can be investigated by the theorem of Routh-Hurwitz's criterion \cite{Merkin,Edmund}. The linearized
matrix of Eq.(\ref{dyn}) around the steady states gives four inequalities to determined
the stable parametric region. As the other two inequalities are cumbersome, only two simple ones are listed below
\begin{eqnarray*}
&&\kappa +\gamma>0, \\
&&1+\kappa ^{2}+\frac{\gamma ^{2}}{4}+\frac{\kappa }{\gamma }%
(\Delta ^{2}+\frac{\kappa ^{2}}{4}+\gamma ^{2})+4\omega _{2}I_{s}>0.
\end{eqnarray*}%
The inequalities imply that a larger $\kappa$ or $\gamma$ can stabilize the system
and validate the static analysis of the parametric effect.
A more detailed discussion about the stability can refer to Ref.\cite{Lin2}.

As the nonlinear responses of the resonator are manifest in a red-detuned pumping
region ($\delta_c<0$) as shown in Fig.\ref{figure1},
in the following sections, we will focus on the nonlinear dynamics of the mechanical
resonator with both linear and quadratic couplings mainly in a red-detuned region.
As the system is not a conservative system without time reversibility
\cite{Lamb}, we only consider the dynamics starting with a zero initial
conditions (ground state) from a control point of view.

\section{Parametric amplification of SSO with step-like amplitudes}

The above nonlinear steady-state responses influenced by the quadratic coupling are based
on the steady-state analysis without considering any dynamical stabilities of the system.
However, when the pumping field is above a threshold power, the static behaviors shown
in Fig.\ref{figure1} will lose their stabilities and run
into a limit-circle motion via a Hopf bifurcation \cite{Okamoto,Zaitsev,Lin}, which is
often identified as the SSO emerged in many nonlinear systems \cite{Jenkins}.
\begin{figure}[htp]
\includegraphics[width=0.24\textwidth]{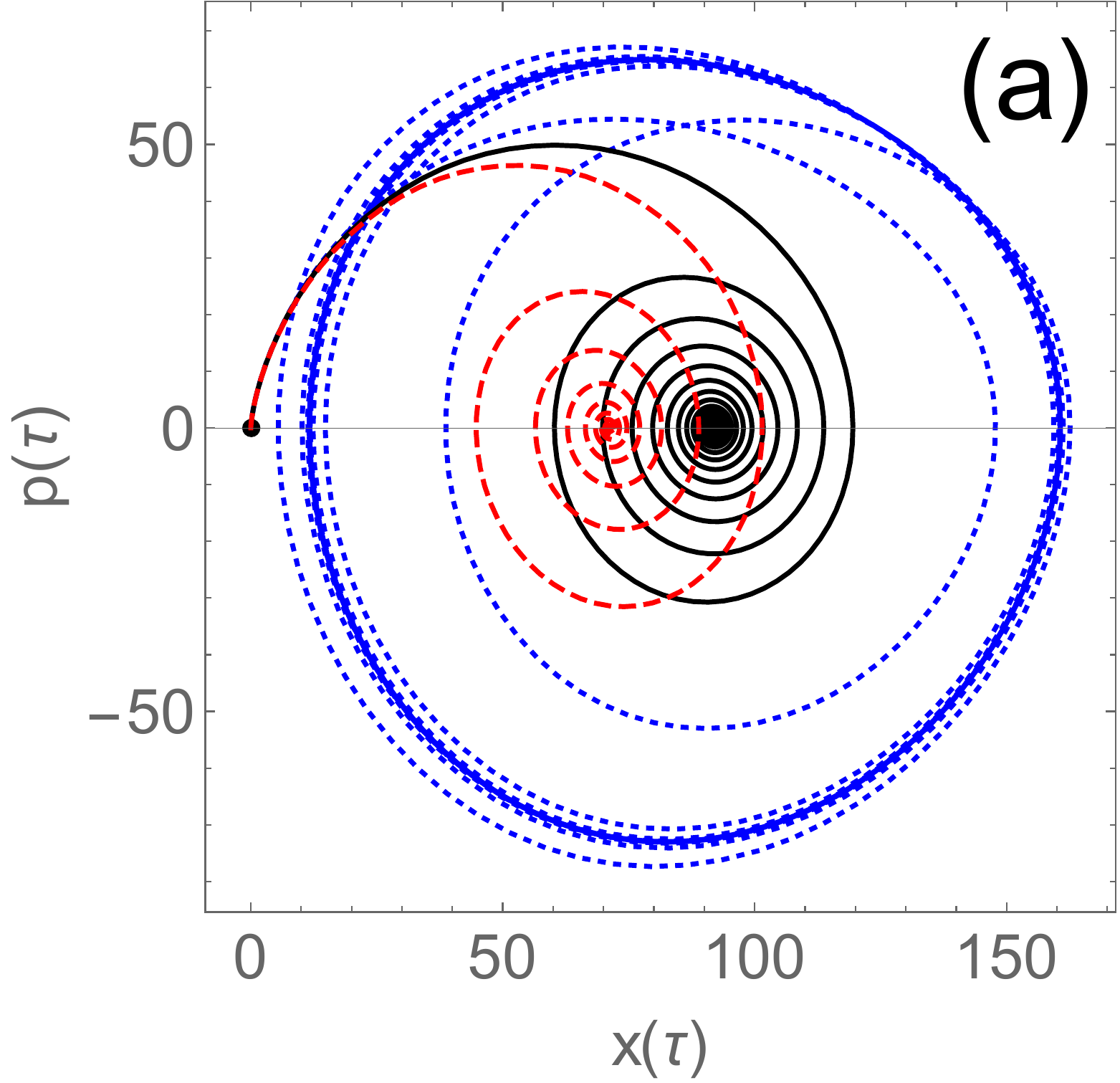}%
\includegraphics[width=0.24\textwidth]{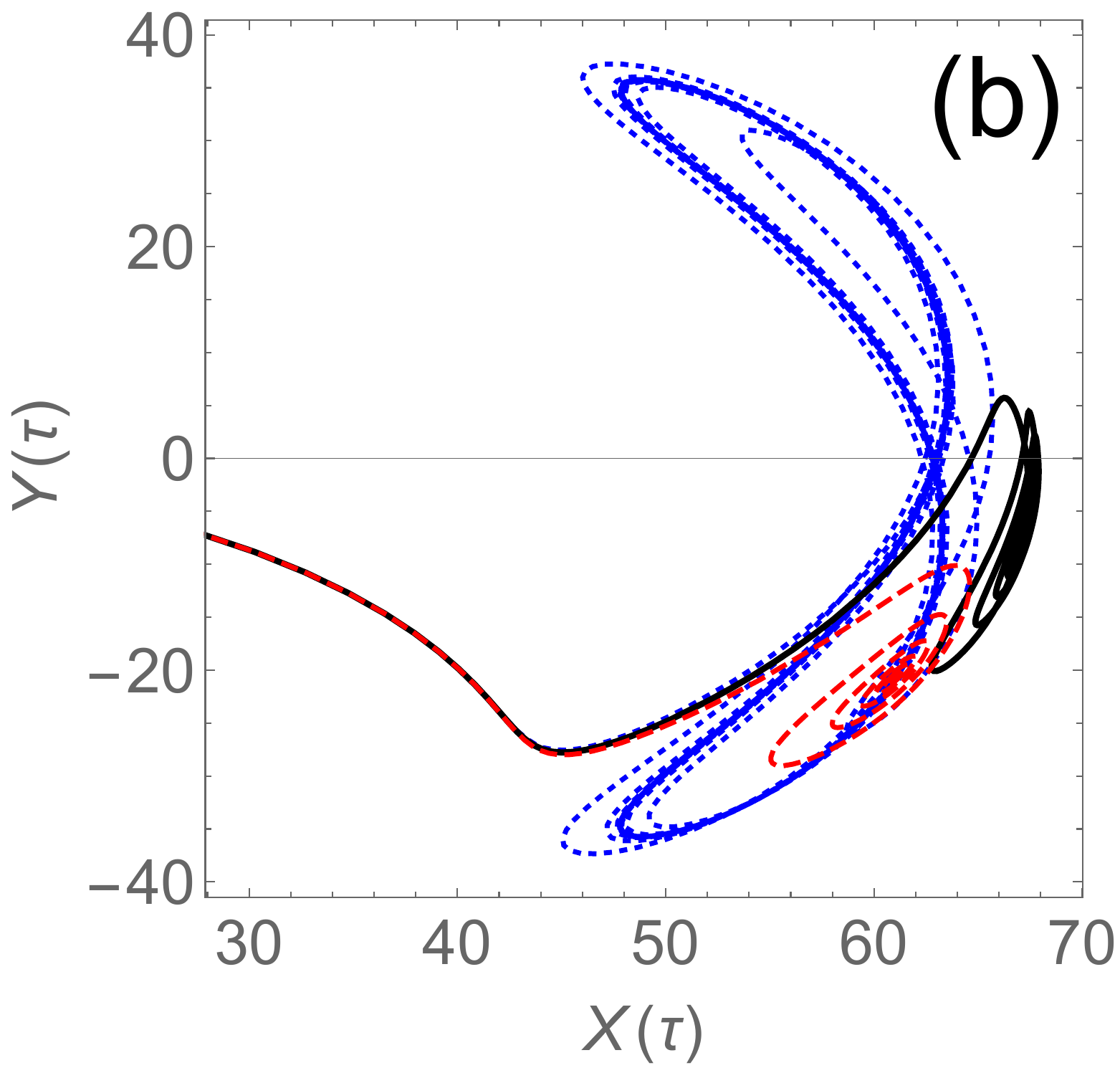}%
\caption{(a) The dynamical trajectories of the mechanical resonator and (b) the
corresponding orbits of the two quadratures of the cavity field in phase space. The parameters
are $\gamma=0.01, \kappa =2$, $\delta_c=-1$, $\eta=68$, $g_{1}=-0.01$ and
the quadratic coupling rates are $g_{2}=0$ (black solid lines), and $g_{2}=-10^{-5}$ (blue dotted lines)
and $g_{2}=10^{-5}$ (red dashed lines), respectively.}
\label{figure3}
\end{figure}%
Fig.\ref{figure3} displays the typical dynamic orbits of the resonator and the
corresponding cavity field in phase space starting from ground states
in a red-detuned pumping field. We can see from Fig.\ref{figure3}(a) that a very
small negative quadratic coupling can sustain a self-oscillation
(blue dotted line) in a parametric region when no SSO exists if only the linear
coupling is included (black solid line). Fig.\ref{figure3}(b) shows that this case
corresponds a very typical parametric amplification process where the frequency of
the cavity field (blue dotted line) is double the frequency of the resonator.
However, a positive quadratic coupling will suppress the transition to SSO under
the same conditions (red dashed lines). When the resonator sustains a self-oscillation
with a period time of $T$ by the parametric amplification, its motion can be approximately
described by
\begin{equation}
\label{SSO}
x(\tau)\approx x_0+A\cos(\Omega \tau),
\end{equation}
where $x_0$ is the mean displacement of the personator mainly induced by the light
pressure force, $A$ is the average amplitude,
and $\Omega=2\pi/T$ is the frequency of the SSO which can be estimated by
a perturbation method with a small displacement as
\begin{equation}
\Omega \approx\sqrt{1+\frac{4g_{2}\eta ^{2}}{\left( \frac{\kappa }{2}\right)
^{2}+\delta _{c}^{2}}-\frac{\gamma _{M}^{2}}{16}}.
\label{Omega}
\end{equation}%
However, for a large-scale displacement, the above frequency will be invalid for the
manifest nonlinear effect.

\begin{figure}[htp]
\includegraphics[width=0.45\textwidth]{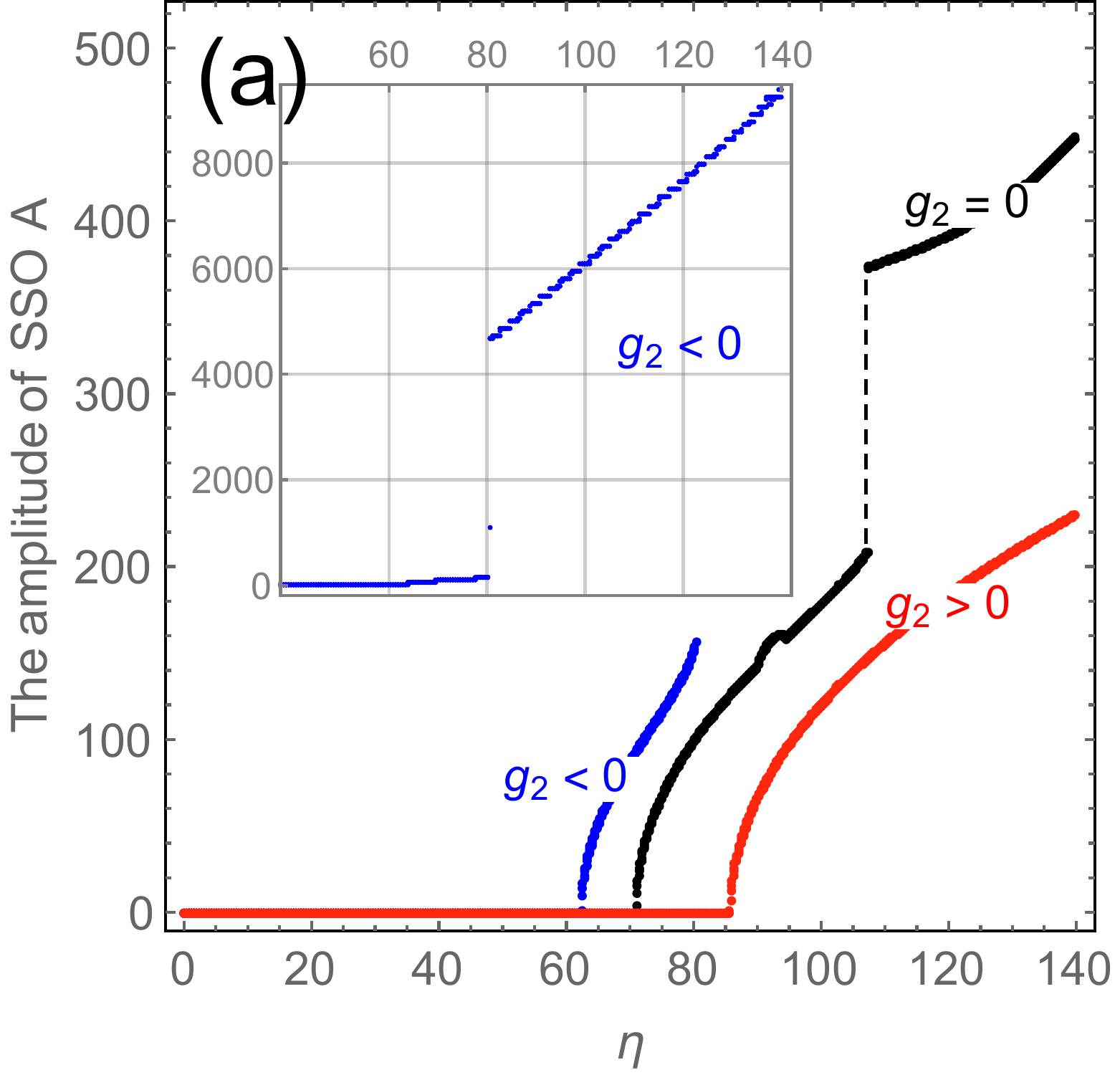}\\%
\includegraphics[width=0.48\textwidth]{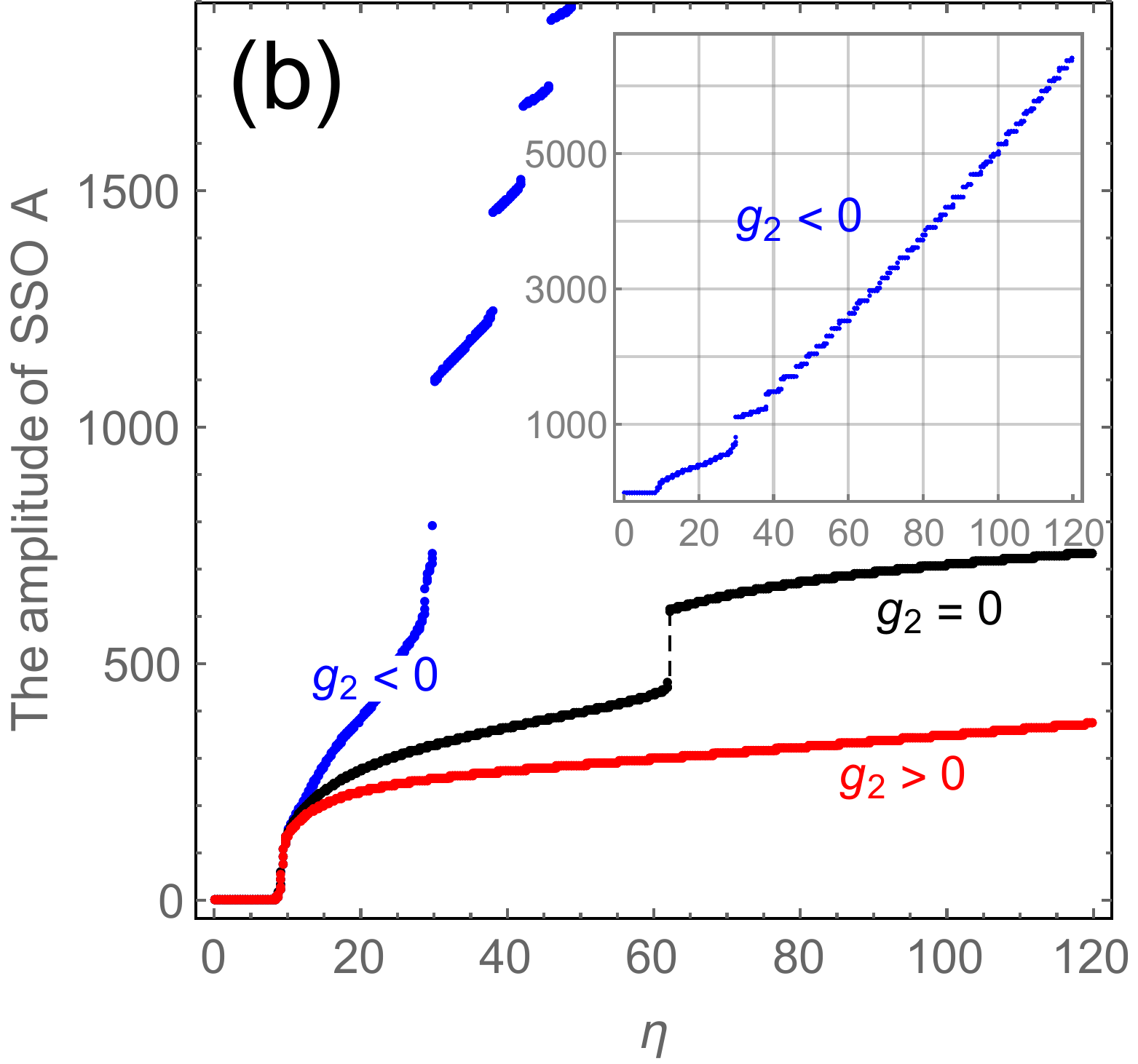}%
\caption{The step-like amplitudes of SSO for
(a) red-detuned case of $\delta_c=-1$ and (b) blue-detuned case of $\delta_c=1$.
Insets: The full view of the amplitude stairs for (a) and (b). The other parameters
are the same as that in Fig.\ref{figure3}.}
\label{figure4}
\end{figure}

The resonator with a quadratic coupling can not only lock easily on SSO
with a controllable amplitude, but also exhibits a very interesting limit-circle behavior
compared with that of a conventional one. Fig.\ref{figure4} demonstrates that a weak
negative quadratic coupling ($g_2=-10^{-5}$) can sustain stable SSOs
with stair-like amplitudes (discrete energies) and a positive $g_2$ can suppress the
step jumping behavior by a smooth increment of amplitude with the same pumping power.
The threshold power bursting into SSO shown in Fig.\ref{figure4}(a) reveals that
the negative quadratic coupling can support SSO (the blue line)
with lower power threshold than that does for a conventional one (the red line) in a red-detuned pumping case,
while, for the blue-detuned case shown in Fig.\ref{figure4}(b), the bifurcation point remains the same with respect
to the quadratic couplings but the amplitude of SSO is dramatically reduced.
The energy balance of loss and gain for the resonator can be used to determine the Hopf bifurcation
points for the self-oscillation under different pumping powers. The parametric effect of the
quadratic coupling on the resonator's dynamic reveals that SSO with discrete amplitudes
can be easily controlled by the quadratic coupling, and the pumping power to sustain a stable self-oscillation
is reduced by a negative quadratic coupling in the red-detuned pumping case as shown in Fig.\ref{figure4}(a).

\begin{figure}[t]
\includegraphics[width=0.45\textwidth]{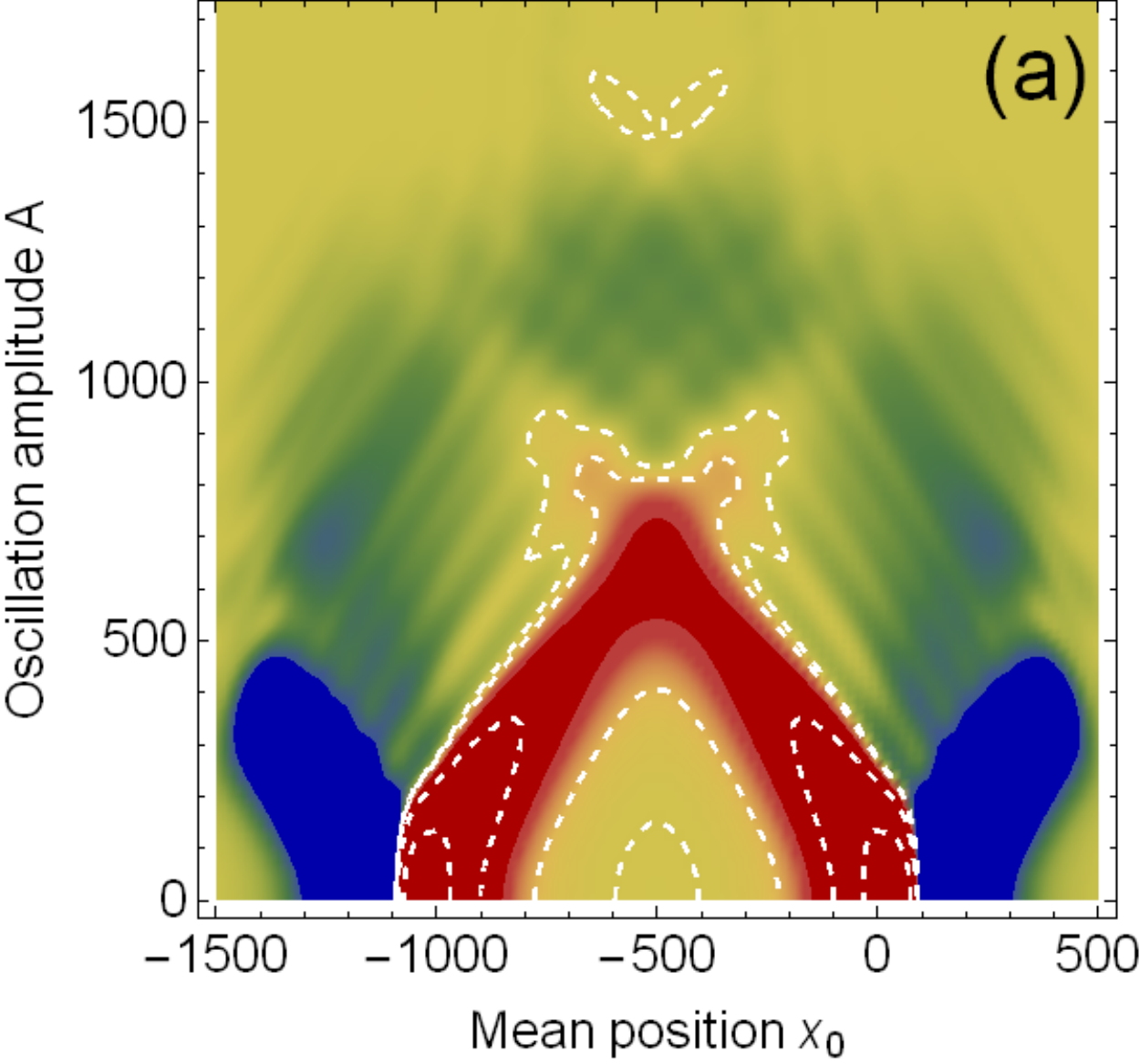}\\%
\includegraphics[width=0.45\textwidth]{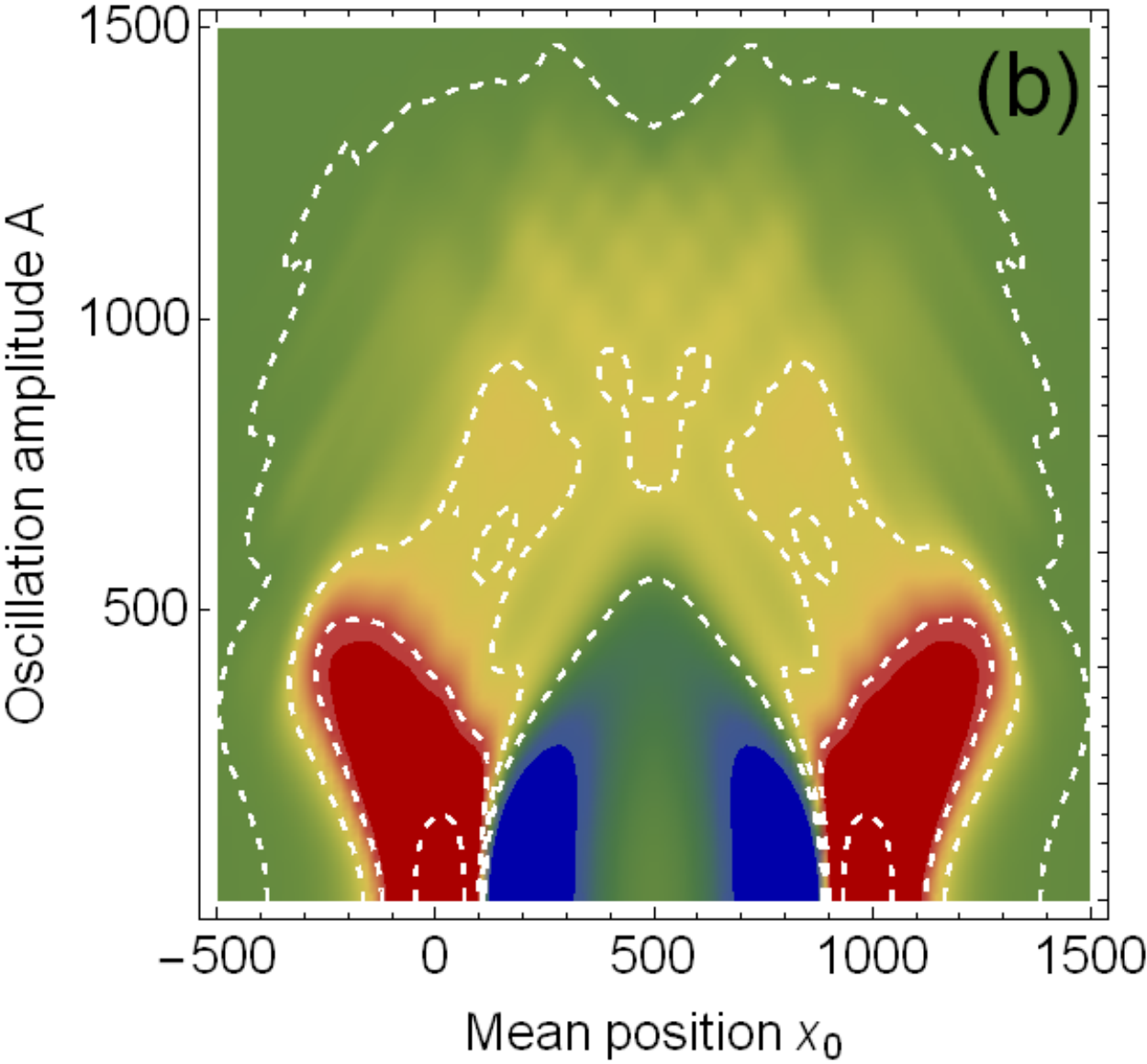}%
\caption{The power balance structure of the resonator for (a) the negative quadratic
coupling of $g_2=-10^{-5}$ and (b) the positive quadratic coupling of $g_2=10^{-5}$ with a
reference pumping amplitude of $\eta=100$.
The white dashed lines are the energy balanced curves calculated in different pumping power of $\eta^2$.
The other parameters are $\delta_c=-1, \kappa =2, g_{1}=-0.01$
and $\Omega$ is determined by Eq.(\ref{Omega}).}
\label{figure5}
\end{figure}

In order to understand the discrete amplitude of the self-oscillation, we can use the adiabatic equation
for the resonator to give a rough analysis. When the resonator locks on a self-oscillation,
the field intensity must also be periodic and the resonator approximately follows the adiabatic equation of
\begin{equation}
\ddot{x}+\frac{\gamma _{M}}{2}\dot{x}+\left( 1+4g_{2}I_c
\right) x=-2g_{1}I_c,%
\label{adabatic}
\end{equation}
where the periodic cavity field $I_c$ is
\begin{equation*}
I_{c}\left( \tau \right)\approx\frac{\eta ^{2}\left( \tau \right) }{\left( \frac{%
\kappa }{2}\right) ^{2}+\Delta ^{2}\left( \tau\right) }.
\end{equation*}
As Eq.(\ref{adabatic}) should have a consistent solution with a form of Eq.(\ref{SSO}),
the motion-modulated detuning reads%
\begin{equation*}
\Delta \left(\tau \right) =A_{0}+A_{1}\cos \Omega t+2 A_{2}\cos 2\Omega t,
\end{equation*}%
where
$A_{0}=(\delta _{c}-g_{1}x_{0}-g_{2}x_{0}^{2})-g_{2}A^{2}/2$,
$A_{1}=-A( g_{1}+2g_{2}x_{0})$, $A_{2} =-g_{2}A^{2}/4$.
Now, $I_c(\tau)$ is a periodic function of time and Eq.(\ref{adabatic}) becomes
a damped and driven Mathieu equation which is stable
for a solution of Eq.(\ref{SSO}) only with the specific parameters. For a constant
pumping power, the Floquet exponents of Eq.(\ref{adabatic}) can be used to analyze
dynamical stability. A simple picture of the discrete amplitude shown in Fig.\ref{figure4} is that the
adiabatic potential of Eq.(\ref{U}) can induce an average multiple-well potential of
\begin{equation*}
\bar{U}\left( A\right) =\frac{1}{T}\int_{0}^{T}U\left( x_0+ A \cos\Omega \tau,\tau \right) d\tau,
\end{equation*}
which supports the self-oscillation with step-like amplitudes with different $x_0$ determined by the
pumping power. A detailed analysis of the discrete amplitudes in this case should resort
to the power balance of $\left\langle\ddot{x}\dot{x}\right\rangle=0$ for the resonator \cite{Girvin},
which gives \cite{Lin}
\begin{equation}
\label{amp}
A\approx\frac{8\left( g_{1}+2g_{2}x_{0}\right) \bar{I}_{1}}{
\Omega \gamma _{M}-8g_{2}\bar{I}_{2} },
\end{equation}
where $\bar{I}_{1,2}$ are the average Fourier sine transforms of the cavity field (see Appendix \ref{app1}).
Eq.(\ref{amp}) reveals that the amplitude of
the self-oscillation can be increased or decreased by the quadratic coupling $g_2$
in the denominator or in the numerator of Eq.(\ref{amp}) due to its parametric effect,
and the resonant parametric amplification of SSO happens at $\bar{I}_2=\Omega \gamma_M/8g_2$.
By using the long-time integral of Eq.(\ref{dyn}) with Eq.(\ref{SSO}), we have (see Appendix \ref{app1})
\begin{equation}
\bar{I}_{1,2}\approx\eta ^{2}\times \mathrm{Im}\left[ \sum_{n,m=-\infty }^{\infty
}\sum_{n^{\prime },m^{\prime }=-\infty }^{\infty }\alpha _{n,m}\alpha
_{n^{\prime },m^{\prime }}^{\ast }\right],
\label{I12}%
\end{equation}%
where $\mathrm{Im}[\cdot]$ means the imaginary part in the square brackets and the sums for
$\bar{I}_{1}$ and $\bar{I}_{2}$ are taken under the resonant
conditions of%
\begin{equation*}
n^{\prime }-n+2\left(m^{\prime}-m\right) +1=0,
\end{equation*}%
and
\begin{equation*}
n^{\prime }-n+2\left( m^{\prime }-m\right) +2=0,
\end{equation*}%
respectively. The parameters $\alpha _{n,m}$ are defined by \cite{Lin}%
\begin{equation*}
\alpha _{n,m}=\frac{J_{n}\left( A_{1}\right) J_{m}\left( A_{2}\right) }{%
\frac{\kappa }{2}-i\left( A_{0}+n\Omega+2m\Omega\right) },
\end{equation*}%
where $J_{n}\left( \cdot \right) $ is the $n$-th order Bessel functions of
the first kind.
In Fig.\ref{figure5} we give a numerical calculation on the power structure of the
resonator determined by
\begin{equation*}
\bar{P}_m(x_0,A)=\frac{g_{1}+2g_{2}x_{0}}{\Omega ^{2}A}\bar{I}_{1}+\frac{g
_{2}}{\Omega ^{2}}\bar{I}_{2}-\frac{\gamma _{M}}{8\Omega }.
\end{equation*}
We can see that the energy ``gain" (the red area $\bar{P}_m >0$) and ``loss" (the blue area $\bar{P}_m <0$) of the resonator
depends on both mean displacement $x_0$ and oscillation amplitude $A$, and the balanced curves
of $\bar{P}_m=0$ for differen pumping powers are shown by the white dashed lines in Fig.\ref{figure5}.
Basically, the light pressure of the cavity field ``pushes" the resonator to give a mean displacement
which changes with the pumping power of $\eta^2$ (see Fig.\ref{figure1}(b)) and the amplitude of the
resonator will modify itself along the white dashed lines and conducts a jump
between different white lines in order to follow the energy balance condition for different pumping powers.
We can identify a clear change of the energy balance structure of the resonator that is sensitively
modified by the quadratic coupling shifting from negative to positive. Fig.\ref{figure5} also indicates
a localized gain region of the resonator (red area) due to the parametric stabilization effect
introduced by the quadratic coupling.

\section{Chaotic dynamics and the suppression of chaotic bifurcations}

\begin{figure}[htp]
\includegraphics[width=0.45 \textwidth]{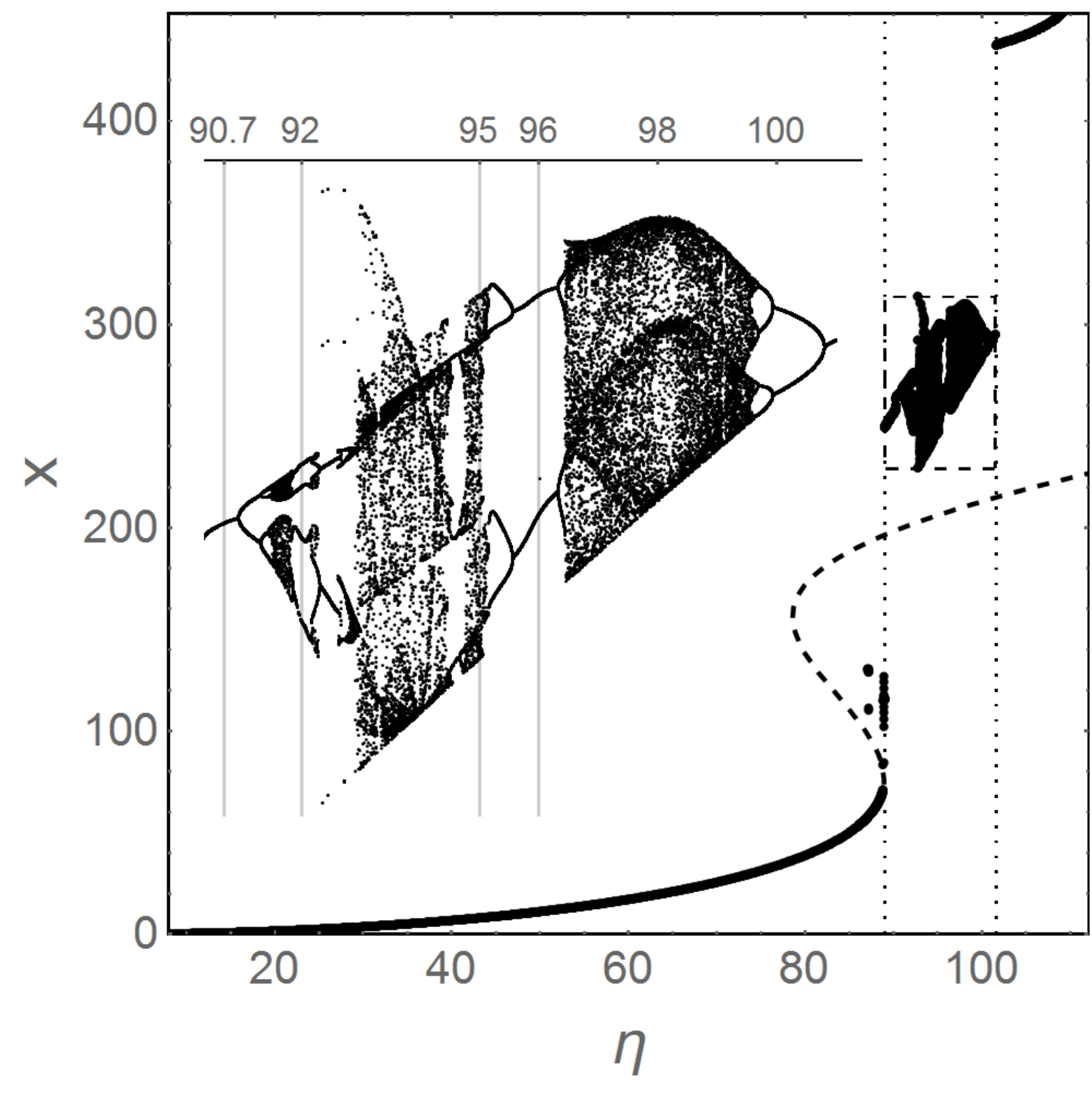}%
\caption{A typical positive bifurcation structure of mechanical resonator's position versus
pumping amplitude $\eta$. The parameters are $\kappa=2,
\gamma_M=0.01, \delta_c=-2, g_1=-0.01$ and $g_2=-10^{-5}$ with the zero initial conditions
of $x(0)=p(0)=X(0)=Y(0)=0$. The thick dashed line is
the steady-state responsive curve. The inset gives the zoomed-in picture from the dashed rectangle.}
\label{figure6}
\end{figure}

However, in a pumping field with an even higher power, the regular self-oscillation will be
unstable and the nonlinear resonator will break into chaos under some parametric conditions.
A typical dynamic transition from a steady-state
(the dashed line determined by Eq.(\ref{stable1})(\ref{stable2})) to chaotic dynamics
is shown in Fig.\ref{figure6} by the jumping and period-doubling bifurcation process.
As shown in Fig.\ref{figure6}, when the input energy from the pumping field
exceeds the energy loss to the environments, the steady states of the resonator bursts into self-oscillations
and then transit to chaotic oscillations
via the typical period-doubling bifurcation routes \cite{Bakemeier,Carmon}.
A complete dynamic behavior of a light-driven resonator shown in Fig.\ref{figure6}
demonstrates three different transitions from steady state to SSO,
to chaotic oscillation and finally back to SSO again along with the increasing power of the pumping
field. The inset of Fig.\ref{figure6} displays the details of the
period-doubling routes to chaos and the inverse bifurcations back to
SSO with respect to the pumping power.
We find a rich structure of the bifurcation-to-inverse-bifurcation window \cite{Girvin}
embedded in the driving parametric region from $\eta=90$ to $\eta=101$ under the parameters
shown in Fig.\ref{figure6}.
\begin{figure}[t]
\includegraphics[width=0.48 \textwidth]{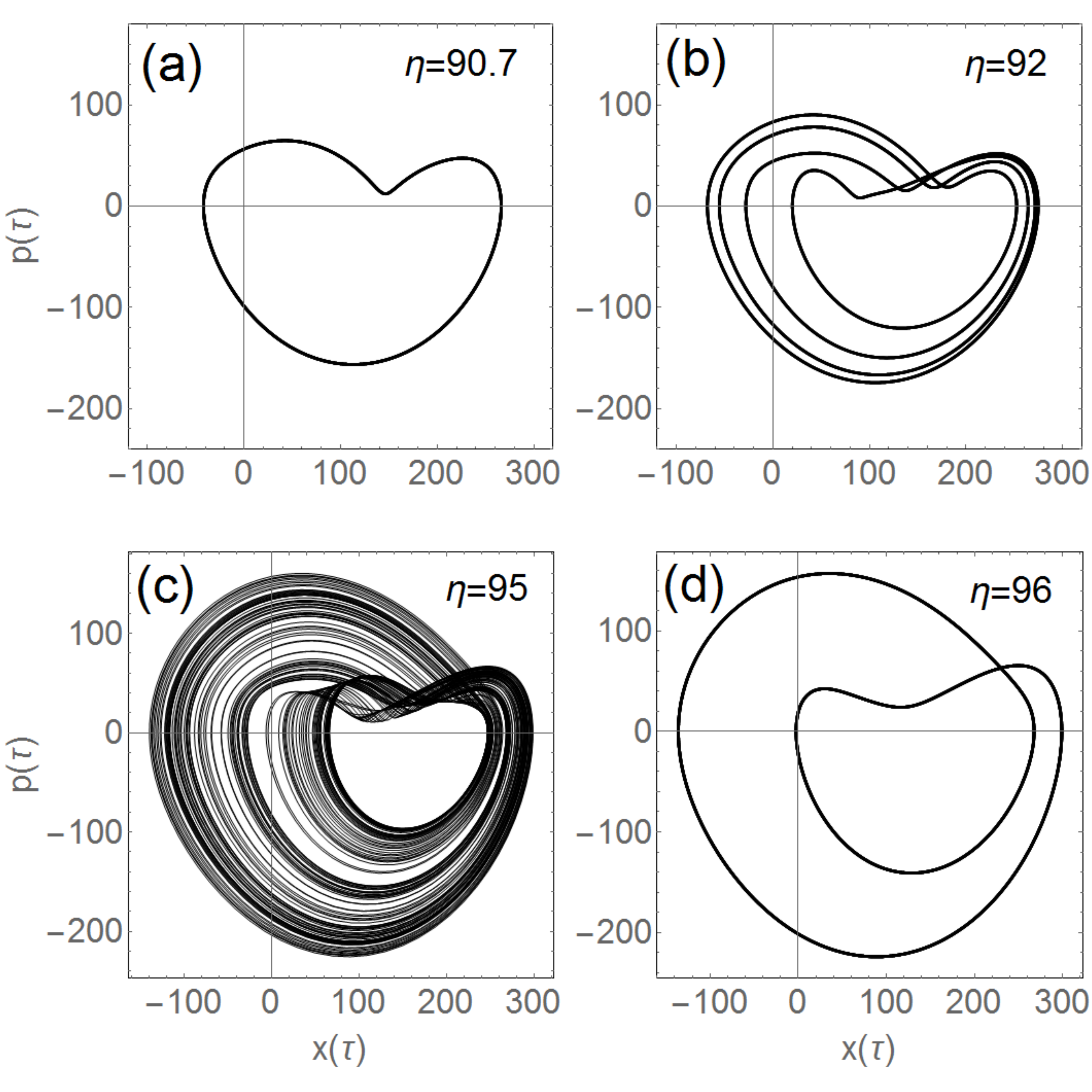}%
\caption{The dynamic orbits of the mechanical resonator in
the phase space for the specific pumping points at (a) $\eta=90.7$, (b) $\eta=92$, (c)
$\eta=95$ and (d) $\eta=96$, respectively. The other parameters are the same
as that in Fig.\ref{figure6}.}
\label{figure7}
\end{figure}%

Fig.\ref{figure7} picks several typical dynamical orbits of the resonator
in the phase space to display one of the bifurcation-to-inverse-bifurcation window.
At $\eta=90.7$ shown in Fig.\ref{figure7}(a), the
resonator takes an attractive period-1 orbit (a stable limit circle)
and then breaks into a period-4 orbit at $\eta=92$ shown in Fig.\ref{figure7}(b). One
chaotic attractor is formed at $\eta=95$ shown in Fig.\ref{figure7}(c) and then returns
back to a 2-period orbit when the pumping amplitude increases to $\eta=96$ through an inverse
bifurcation process.
\begin{figure}[htp]
\includegraphics[width=0.48 \textwidth]{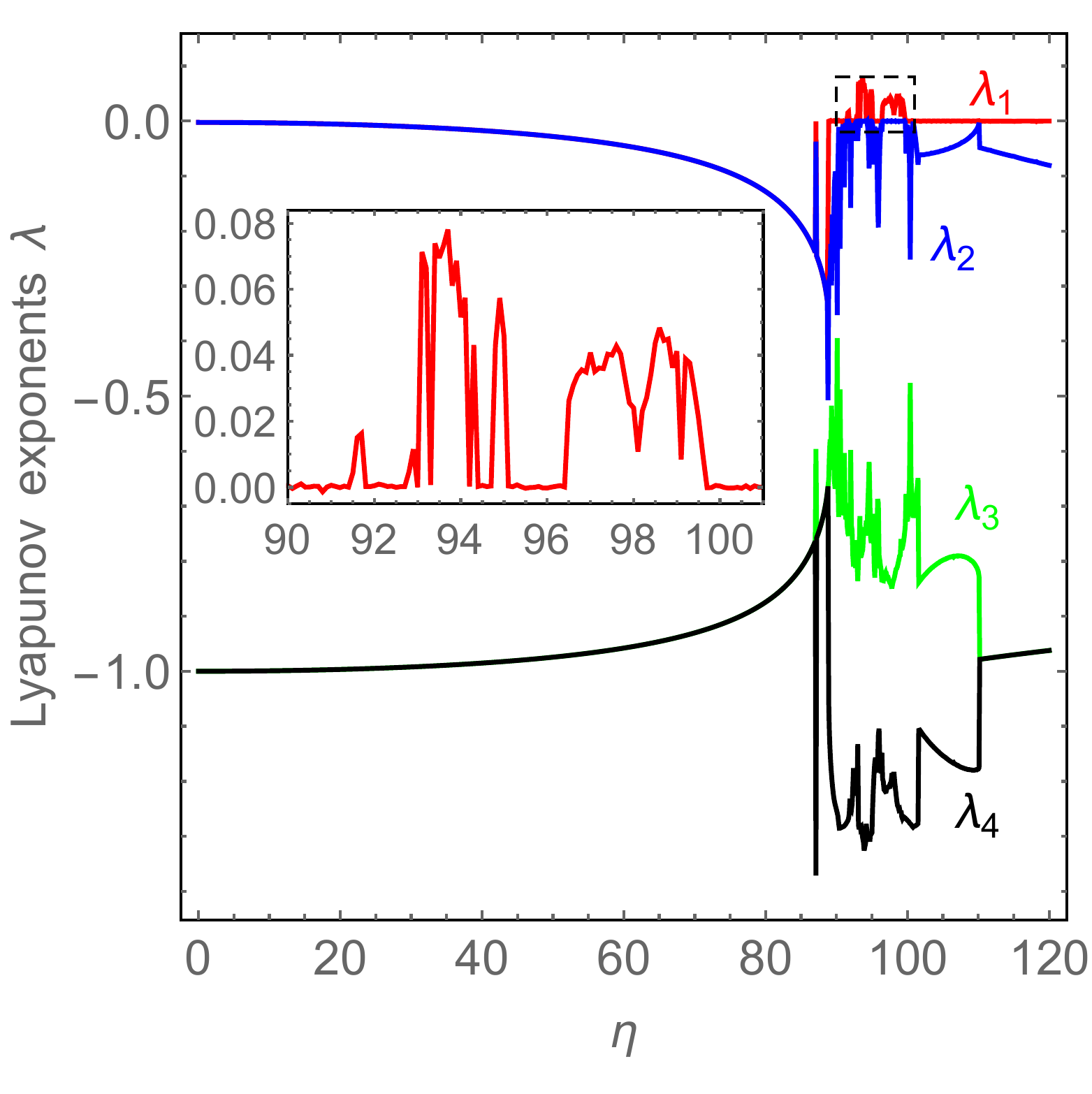}\\%
\caption{The Lyapunov exponents $\lambda_1, \lambda_2, \lambda_3$ and $\lambda_4$
of the optomechanical system corresponding to Fig.\ref{figure6}.
The inset shows the zoomed-in image from the
dashed rectangle of the maximal Lyapunov exponent $\lambda_1$.
All the parameters are the same as that in Fig.\ref{figure6}.}
\label{figure8}
\end{figure}

We can identify this interesting chaotic dynamics by calculating the
Lyapunov exponents (LEs) of the system as shown in Fig.\ref{figure8}.
We can clearly see that the maximal Lyapunov exponent (MLE) can
be used to discriminate three different dynamics, the steady-state
behavior with a negative MLE, the limit-circle motion with a
zero MLE, and a positive MLE in the chaotic window along
with abrupt droppings near to zero (inset of Fig.\ref{figure8}). The very
interesting thing is that the stability of the system is enhanced by
a clear dropping of MLE at the critical pumping threshold around $\eta=89$.
This effect is due to the static ``spring'' effect of the light field \cite{Zadeh,Sheard}
which is proportional to $\eta^2$ (see Eq.(\ref{UL})) and can also be modified by the quadratic
coupling. The zero value of MLE outside the chaotic window is a clear signature of the limit-circle
dynamics of the resonator. For a motion on a limit circle, the LEs of the motion perpendicular to
the orbit are negative but the LE along the trajectory is zero indicating a phase freedom of the
limit-circle motion. This phase freedom along the orbit is the well-known mechanism leading to
synchronization of a collection of resonators because the resonators can freely adjust their
relative phases to achieve a synchronized motion \cite{Pikovsky}.
\begin{figure}[htp]
\includegraphics[width=0.48 \textwidth]{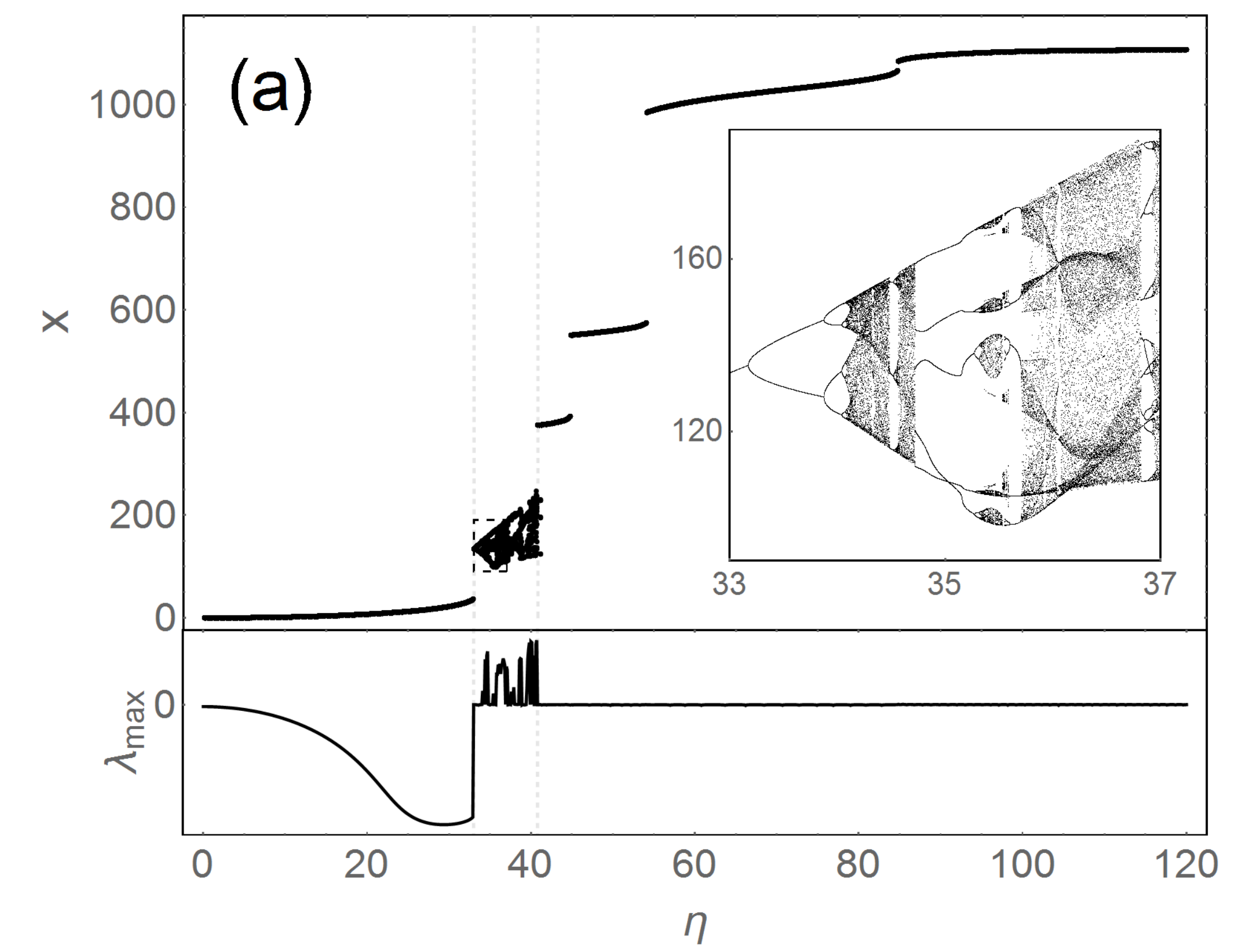}\\%
\includegraphics[width=0.48 \textwidth]{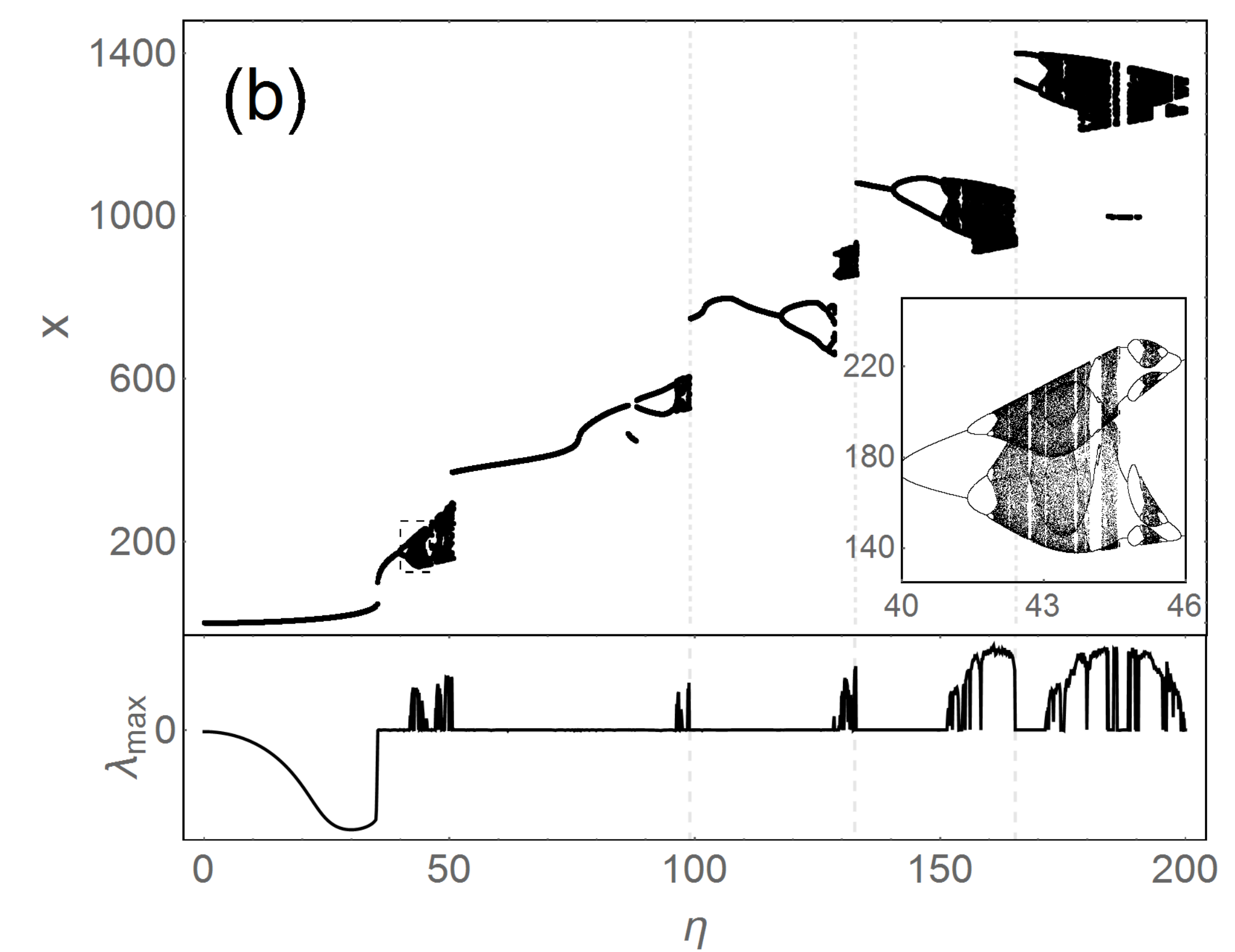}\\%
\includegraphics[width=0.48 \textwidth]{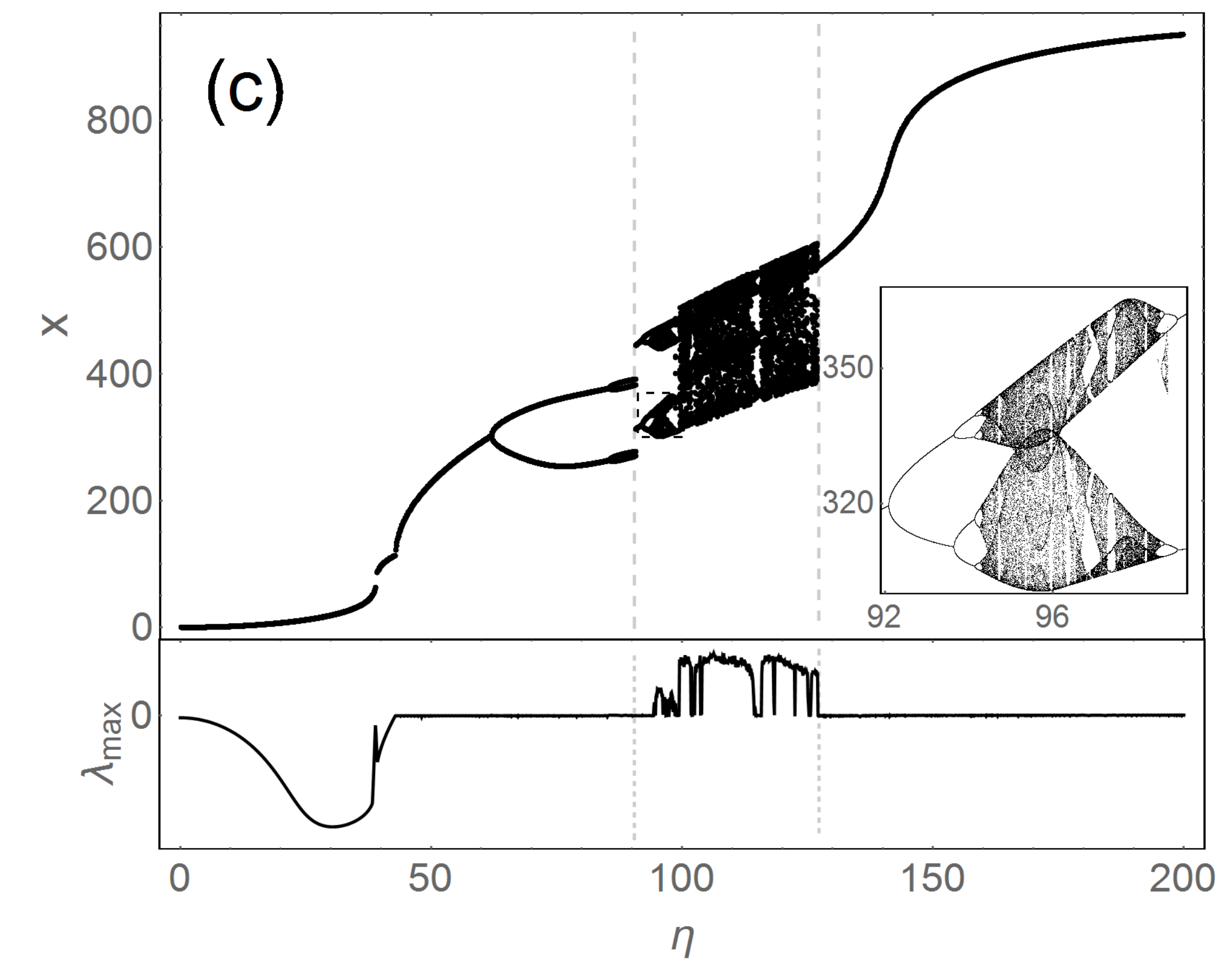}\\%
\caption{Positive displacement bifurcations (top) and the corresponding maximal Lyapunov
exponents (bottom) of the resonator vs
pumping amplitude $\eta$ modified by the quadratic couplings of (a) $g_{2}=-10^{-5}$,
(b) $g_{2}=0$, and (c) $g_{2}=10^{-5}$, respectively. The insets show the zoomed-in bifurcation images
from the corresponding dashed rectangles. The other parameters are
$g_{1}=-0.01, \kappa =1, \gamma_M=0.01, \delta_c=-1$ with zero initial conditions.}
\label{figure9}
\end{figure}

The chaotic dynamics of the mechanical resonator has been found extensively
in the conventional optomechanical systems \cite{Girvin,Carmon,Bakemeier,Larson,Ying},
but, in a system with both linear and quadratic couplings, the bifurcation
and the inverse bifurcation with respect to light driving
can be easily modified by a weak quadratic coupling.
The sensitive modifications on chaotic transition in a nearly resolved sideband
case ($\kappa=\omega_M$) are shown in Fig.\ref{figure9}. The
different dynamics under different quadratic coupling rates clearly verify the parametric modulating effect
of the dynamics on the mechanical resonator, which, totally, stabilizes the motion of the resonator
(compare Fig.\ref{figure9}(a)(c) with Fig.\ref{figure9}(b)) \cite{Seok}.
In a weak pumping power, the mechanical resonator damps
to steady state with a
negative MLE due to the overwhelming energy dissipations and its nonlinear dynamics
can be approximately explained by the adiabatic motion under potential of Eq.(\ref{U}) \cite{Zaitsev}.
With an increasing pumping power, the damping of the oscillation will be balanced and a stable SSO
is established \cite{Lin,Ludwig}, which is indicated by a zero MLE
shown at the bottom frames of Fig.\ref{figure9}. The interesting thing is that the resonator
will adjust its position with a process of bifurcation and inverse bifurcation, and then follows by a
jump to a larger position with a new bifurcation. The MLE shown in Fig.\ref{figure8}
indicates that there exist several bifurcation windows at different pumping powers and one
of which is picked to show by the insets of Fig.\ref{figure9}. When the pumping power becomes stronger,
the mechanical resonator will lock itself to a stable oscillation governed by the light pressure
of $-2g_1 I(\tau)$. In this extreme case, the resonator will be approximately
treated as a forced resonator because the right term of Eq.(\ref{cdeq}) will dominate the resonator's dynamics.

\section{Field signature of dynamic transitions}

As the intensity of the cavity field parametrically couples to the motion of
mechanical resonator, the dynamical transition of the mechanical resonator
can definitely influence the cavity field through a back-action effect. If we stabilize the pumping
power of the driving field, then the longtime amplitude of the cavity field
is given by (see Appendix \ref{app1})%
\begin{equation}
a_{c}\left( \tau \right) \approx \eta e^{-\frac{\kappa }{2}\tau +i\theta
\left( \tau \right) }\int_{0}^{\tau }e^{\frac{\kappa }{2}t-i\theta \left(
t\right) }dt,  \label{ac}
\end{equation}%
where the field frequency of $a_c(\tau)$ is now determined by the position-dependent integral of%
\begin{equation}
\theta \left( \tau \right)=\delta _{c}^{\prime }\tau%
-g_{2}\int_{0}^{\tau }x^{\prime 2}\left(t\right) dt,  \label{freq}
\end{equation}%
where $\delta _{c}^{\prime } =\delta _{c}+g_{1}^{2}/4g_{2}$,
$x^{\prime}(t)=x(t)+g_{1}/2g_{2}$.
Eq.(\ref{freq}) shows that the spectrum of the cavity field is closely related to
the dynamical phase of $\theta(\tau)$, which depends on the energy of the resonator
estimated by $E(\tau) \propto x^{\prime 2}(t)$. Generally,
for a regular motion, the resonator can be described by $x(t)\approx x_0+\sum_{n}A_n(t)\cos(\Omega_n t)$,
where $A_n(t)$ is the instaneous amplitude of the motion corresponding to the component of $\Omega_n$,
but for a chaotic motion, $A_n(t)$ will be a continuous function of $A(\Omega,t)$. Therefore, a close
connection between the dynamics of resonator and the spectrum of the cavity field can be
established by Eq.(\ref{ac}). Based on the input-output relation of the cavity field, the dynamic transition
of the resonator can be readily detected from the output field of the cavity mode.

\begin{figure}[htp]
\includegraphics[width=0.48 \textwidth]{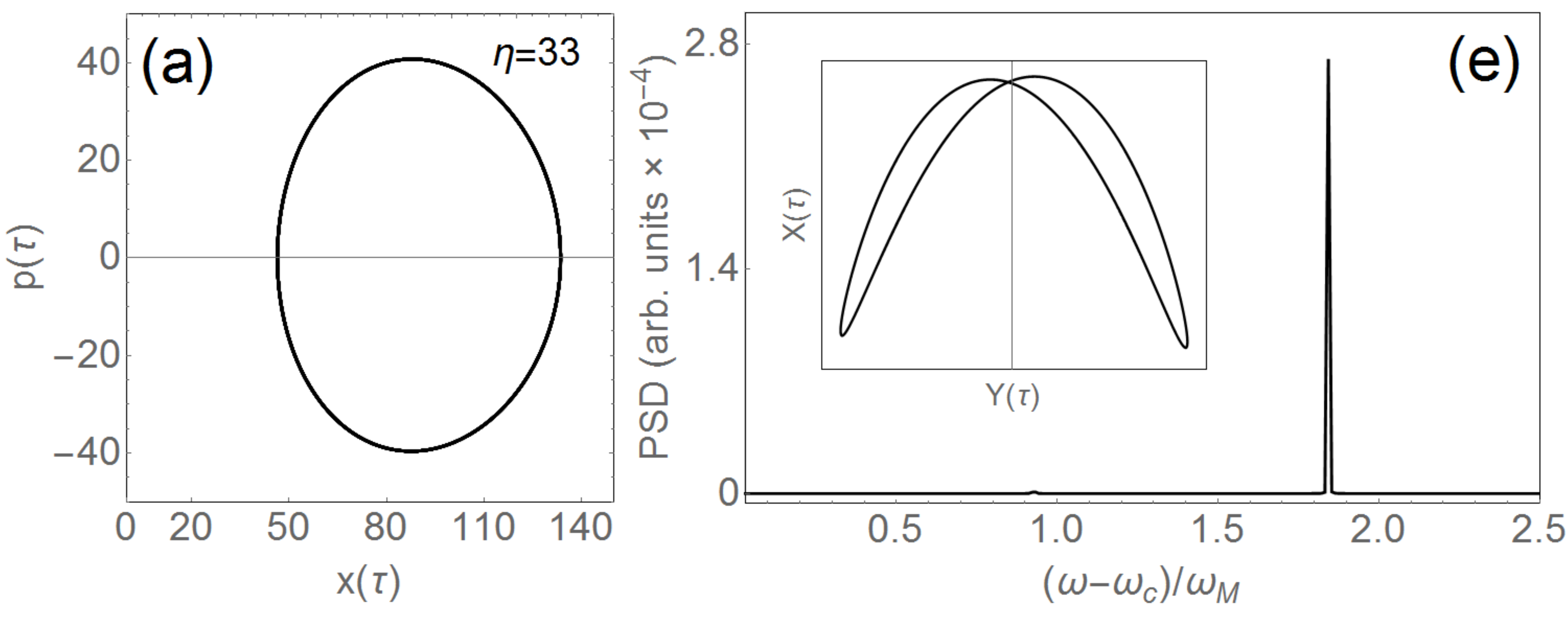}\\%
\includegraphics[width=0.48 \textwidth]{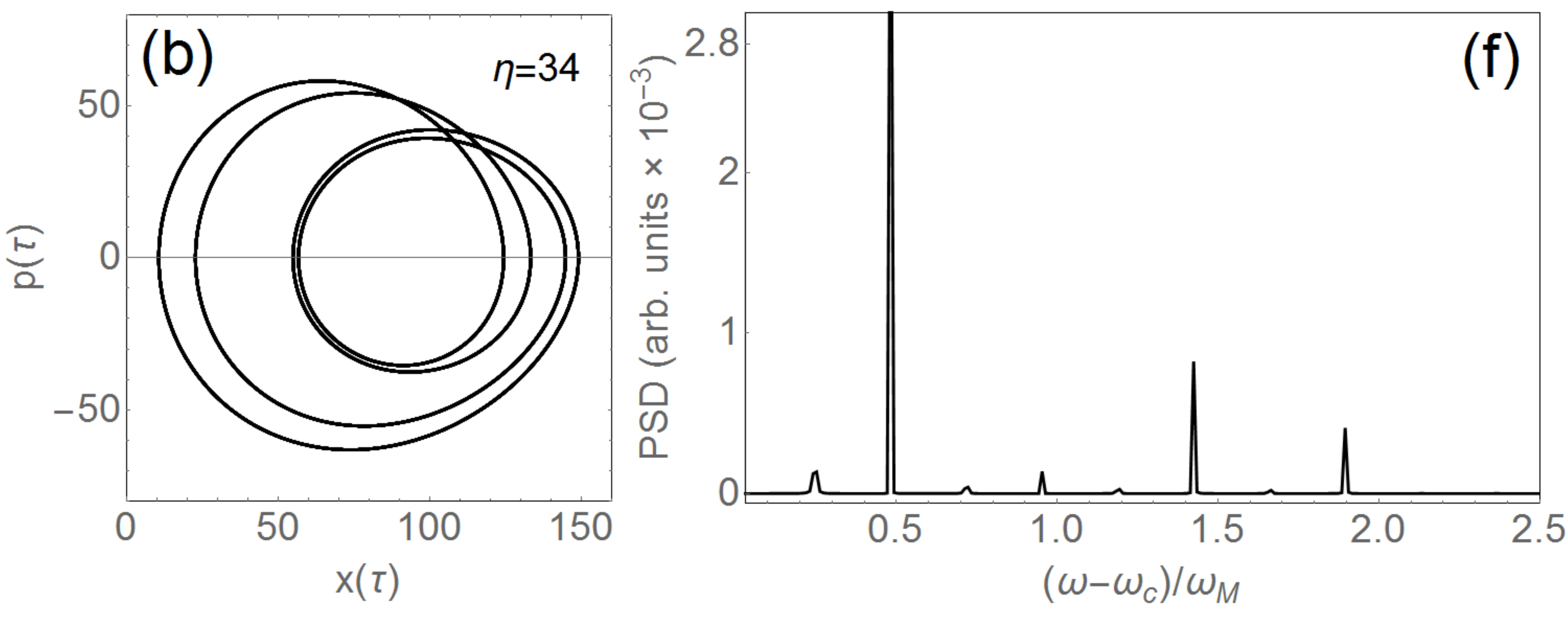}\\%
\includegraphics[width=0.48 \textwidth]{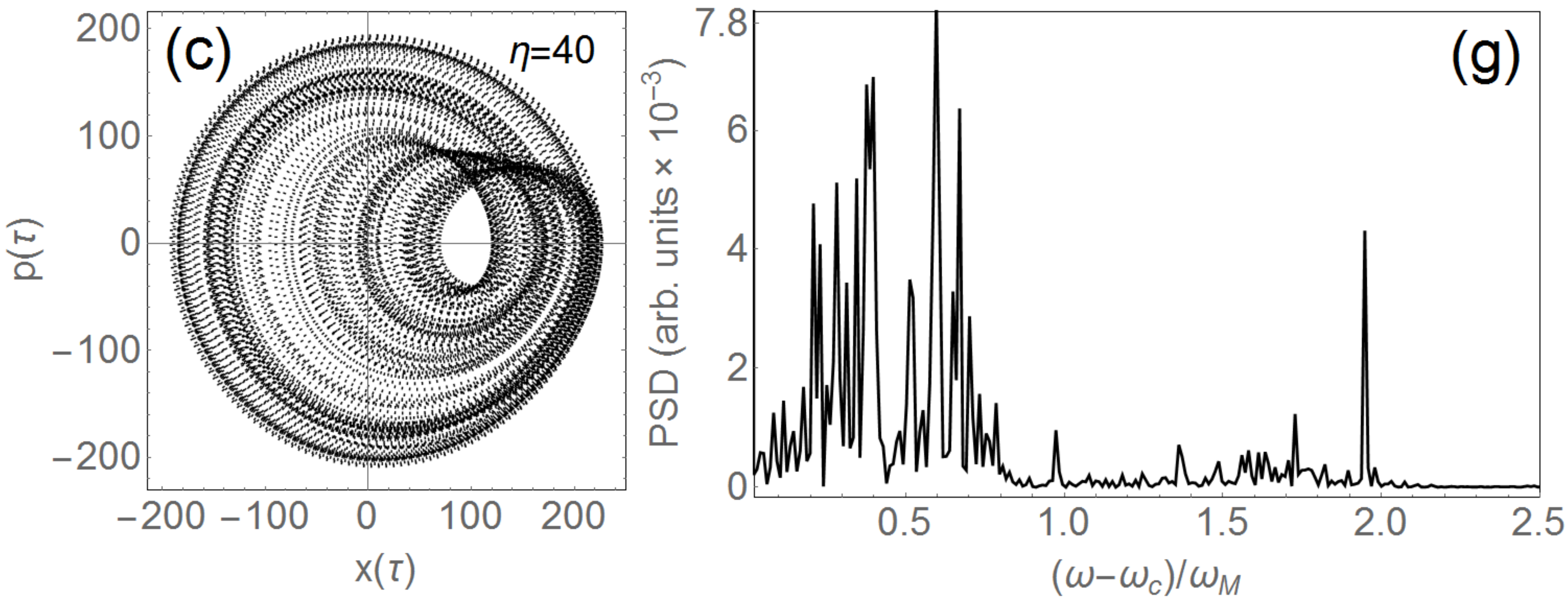}\\%
\includegraphics[width=0.48 \textwidth]{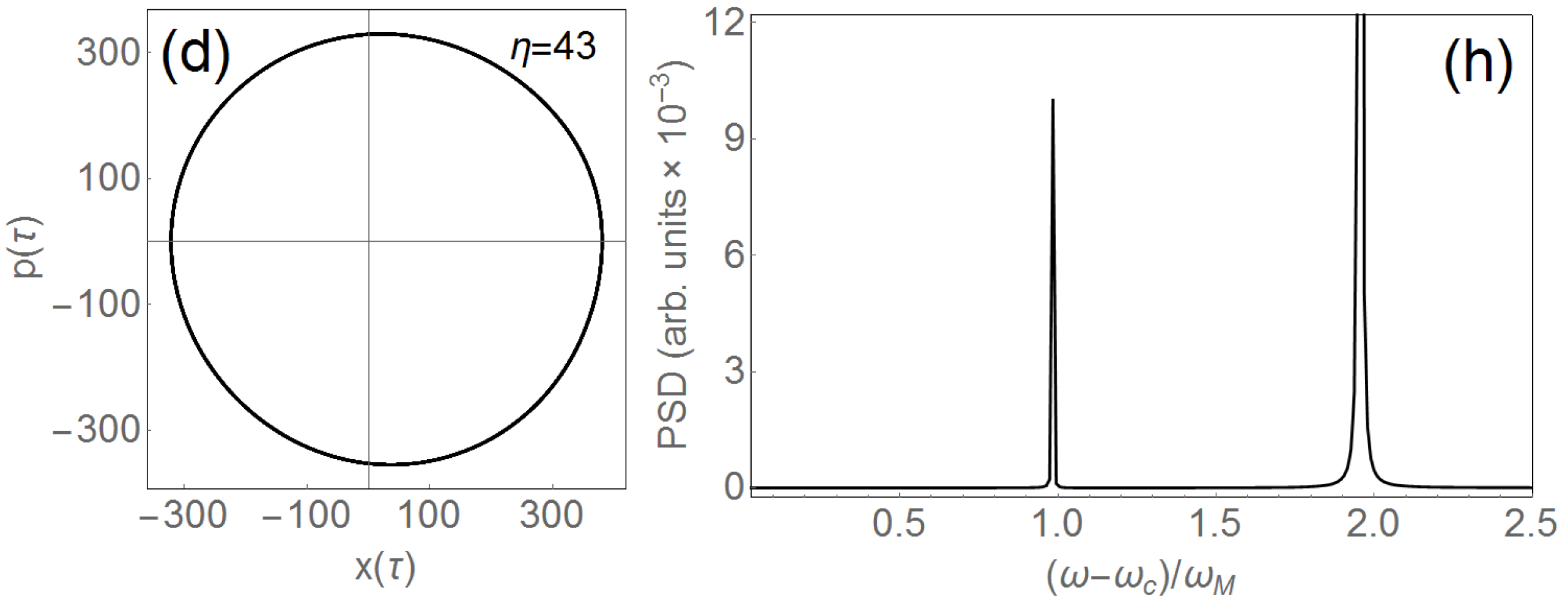}\\%
\caption{The dynamical orbits of the mechanical oscillator in the phase space
(the left column) and the corresponding
power spectral density of the cavity fields (the right column).
The dynamical parameters are $\delta_c=-1$, $\protect\kappa=1$,
$\gamma=0.01$, $g_1=-0.01$ and $g_2=-10^{-5}$.}
\label{figure10}
\end{figure}

By investigating the power spectral density (PSD) of the cavity field,
the motions of the mechanical resonator can be easily traced down. Fig.\ref{figure10}(a)-(d)
present four sample orbits of the mechanical resonator in the phase space with a negative quadratic
coupling rate, and the corresponding power spectra of
the cavity fields are displayed in the right column from (e) to (h).
Fig.\ref{figure10} indicates a close connection of the mechanical dynamics with
the power spectrum of the cavity field. In a relative low pumping
power at $\eta=33$, the resonator maintains its intrinsic frequency on a limit circle
but has a very weak back action on the cavity field. The power spectrum of the cavity field
exhibits a very weak side-band peak near the field-modulated frequency of $\Omega$
(see Eq.(\ref{Omega})) and a relative stronger parametric peak near $2\Omega$. This spectral profile
for a parametric amplification can be verified by the phase trajectory of the field quadratures, $X$ and $Y$, shown by
the inset of Fig.\ref{figure10}(e). When the pumping amplitude increases to $\eta=34$, the resonator
steps into a period-doubling region and the power spectrum exhibits more resonant side-band peaks
at harmonic and subharmonic frequencies around $\omega_c \pm n \Omega/m$, where $n$ and $m$
are integers.
A chaotic dynamics (MLE is positive) at pumping
amplitude of $\eta=40$ will induce a nearly continuous power spectrum for the cavity field as shown in
Fig.\ref{figure10}(g) \cite{Carmon}. When the pumping amplitude reaches $\eta=43$,
the mechanical resonator returns back to a limit-circle dynamics and the corresponding
power spectrum displays a higher peak of the double frequency for an enhanced parametric amplification
 as shown in Fig.\ref{figure10}(h) \cite{Girvin}.
Therefore the power spectrum of the cavity field can be an indicator to trace the dynamical
transition of the mechanical resonator and then a feedback control on the dynamical
transition of the mechanical resonator can be subsequently developed.

\section{Conclusions and Discussion}

By analyzing an optomechanical resonator with both linear and quadratic couplings,
we give an explicit photon-phonon parametric model to find an interesting SSO with step-like amplitude
and a tunable chaotic bifurcation to inverse bifurcation of the resonator
based on the mean-field dynamics of Eq.(\ref{dyn}) in a red-detuned pumping field.
We investigate the parametric effects of the quadratic couplings to modify the dynamics
of mechanical resonator and compare the different dynamical transitions
between limit-circle oscillations and the chaotic dynamics. The static analysis
reveals a very sensitive quadratic effect on the position control of the mechanical
resonator. The dynamical calculations give new features of the quadratic couplings
and find interesting dynamical bifurcations of the mechanical resonator inducing
different backactions on the field spectrum. Our study, firstly, shows that the
optomechanical system with quadratic coupling provides a direct parametric process which can
be actuated and controlled simultaneously by the intensity of the cavity field. This photon-phonon
parametric system has advantages that its nonlinear behavior can be easily controlled by the
pumping field and it can work in a red-detuned region with a low pumping power and thermal noise. The
parametric effect found in this system gives a controllable oscillator to study the dynamical
transitions, chaotic behaviors and the parametric process by a crossover from classical to
quantum regime \cite{Karabalin}. The study of classical squeezing \cite{Rugar,Girvin0}
verified by the parametric effect also presents refined applications based on the nonlinearity of
this system, such as to realize coherent frequency mixing and parametric coupling network
of optomechanical systems with quadratic couplings. The classical dynamics shows the
possibility of squeezing on the resonator by quadratic coupling along with
a cooling by the linear coupling simultaneously \cite{Girvin0}. The dynamics
of this system are much richer than the conventional optomechanical system,
and it can be easily detected by the power spectra of the field mode.

With both the quadratic and linear coupling, the quantum chaotic effects on the ground
state will be an interesting topic of this model \cite{Ying}. The quadratic coupling can
generate non-classical states and give quantum nondemolition measurement of the discrete energies of
the nano-mechanical resonator \cite{Milburn,Vanner}. In quantum regime, the strong
cross-correlation between the light's and the resonator's motion will lead to new dynamical
features beyond the mean-field dynamics. To explore these possibilities, a further work of
nonlinear effect on the full quantum dynamics should be done on this model.

\begin{acknowledgements}
This work is supported by the National Natural Science Foundation of China (NSFC) Grant No.11447025,
No.11234003 and the Scientific Research Foundation for the Returned Overseas Chinese
Scholars, State Education Ministry.
\end{acknowledgements}

\begin{appendix}
\section{The spectrum of the cavity field}
\label{app1}
The amplitude of the cavity field is determined by
\begin{equation}
\frac{d}{d\tau }a_c\left( \tau \right) =\left[ i\Delta \left( x,\tau \right) -%
\frac{\kappa }{2}\right] a_c\left( \tau \right) +\eta \left( \tau \right) ,
\end{equation}%
and the formal solution is%
\begin{eqnarray*}
a_{c}\left( \tau \right)  &=&e^{-\frac{\kappa }{2}\tau }e^{i\int_{0}^{\tau
}\Delta \left( x,t \right) dt} \\
&&\times \left[ a\left( 0\right) +\int_{0}^{\tau }e^{\frac{\kappa }{2}%
t}e^{-i\int_{0}^{t}\Delta \left( x,t^{\prime }\right) dt^{\prime }}\eta
\left( t\right) dt\right].
\end{eqnarray*}%
For a long time limit $\tau\rightarrow \infty$, the first term disappears, and then
\begin{equation}
a_{c}\left( \tau \right) \approx e^{-\frac{\kappa }{2}\tau +i\theta
\left(\tau \right) }\int_{0}^{\tau }e^{\frac{\kappa }{2}t-i\theta \left(
t \right) }\eta\left( t \right) dt,
\end{equation}%
where the dynamical phase integral is defined by%
\begin{equation*}
\theta \left( \tau \right) \equiv \int_{0}^{\tau }\Delta \left( x,t
\right) dt.
\end{equation*}%
If we put Eq.(\ref{SSO}) into the above equation and set $\eta \left( t\right)=\eta $
to be a constant pumping, we have
\begin{eqnarray*}
a_{c}\left( \tau \right)  &\approx&\eta e^{\left( iA_{0}-\frac{\kappa }{2}\right) \tau
}e^{iA_{1}\sin \left( \Omega \tau \right) }e^{iA_{2}\sin \left( 2\Omega \tau
\right) }\\
&&\times \sum_{n,m=-\infty }^{\infty
}J_{n}\left( A_{1}\right) J_{m}\left( A_{2}\right) F\left( \tau \right),
\end{eqnarray*}%
where $e^{-iA\sin \left(\Omega t\right) }=\sum_{n=-\infty }^{\infty
}J_{n}\left( A\right) e^{-in\Omega t}$ and%
\begin{eqnarray*}
F\left( \tau \right)  &=&\int_{0}^{\tau }e^{\left[ \frac{\kappa }{2}-i\left(
A_{0}+n\Omega+2m\Omega\right) \right] t}dt \\
&=&\frac{e^{\left[ \frac{\kappa }{2}-i\left( A_{0}+n\Omega+2m\Omega\right) \right] \tau
}-1}{\frac{\kappa }{2}-i\left( A_{0}+n\Omega+2m\Omega\right) }.
\end{eqnarray*}%
Therefore, the long-time amplitude of the cavity field is
\begin{eqnarray*}
a_{c}\left( \tau \right)  &\approx &\eta e^{iA_{1}\sin \left(\Omega \tau \right)
}e^{iA_{2}\sin \left( 2\Omega\tau \right) } \\
&&\times \sum_{n,m=-\infty }^{\infty }\frac{J_{n}\left( A_{1}\right)
J_{m}\left( A_{2}\right) }{\frac{\kappa }{2}-i\left( A_{0}+n\Omega+2m\Omega\right) }%
e^{-i\left( n+2m\right) \Omega\tau },
\end{eqnarray*}%
and its intensity is
\begin{equation*}
I(\tau )\approx\eta ^{2}\sum_{n,m;n^{\prime },m^{\prime }}\alpha _{n,m}\alpha
_{n^{\prime },m^{\prime }}^{\ast }e^{i\left[ \left( n^{\prime }-n\right)
+2\left( m^{\prime }-m\right) \right] \Omega \tau }.
\end{equation*}
The time-average integrals of $\bar{I}_{1}$ and $\bar{I}_{2}$ are defined by%
\begin{eqnarray*}
\bar{I}_{1} &\approx &\left\langle I(\tau )\sin (\Omega\tau) \right\rangle =\mathrm{Im}\left[
\frac{1}{T}\int_{0}^{T}I(\tau )e^{i\Omega \tau }d\tau \right] , \\
\bar{I}_{2} &\approx &\left\langle I(\tau )\sin (2\Omega \tau) \right\rangle =\mathrm{Im}\left[
\frac{1}{T}\int_{0}^{T}I(\tau )e^{2i\Omega \tau }d\tau \right],
\end{eqnarray*}%
where $T=2\pi/\Omega$.
If we use the formula%
\begin{equation}
\frac{\Omega}{2\pi}\int_{0}^{2\pi/\Omega}e^{i\left( n^{\prime }-n\right) \Omega \tau }d\tau
=\delta _{n,n^{\prime }},
\end{equation}%
we can obtain Eq.(\ref{I12}) with the corresponding sum rules.
\end{appendix}


\begin{thebibliography}{99}
\bibitem{Marqu} M. Aspelmeyer, T. J. Kippenberg, and F. Marquardt,
\textit{Cavity optomechanics},
Rev. Mod. Phys. \textbf{86}, 1391 (2014).

\bibitem{Clerk} A. A. Clerk, M. H. Devoret, S. M. Girvin, F. Marquardt,
and R. J. Schoelkopf. \textit{Introduction to quantum
noise, measurement, and amplification}, Rev. Mod. Phys.
\textbf{82}, 1155 (2010).

\bibitem{Poot} M. Poot and H. S. J. vander Zant, \textit{Mechanical systems in the quantum regime},
Phys. Rep. \textbf{511}, 273 (2012).

\bibitem{Meystre} M. Aspelmeyer, P. Meystre and K. Schwab,
\textit{Quantum optomechanics}, Phys. Today \textbf{65},
29 (2012).
\bibitem{Thompson} J. D. Thompson, B. M. Zwickl, A. M. Jayich, F. Marquardt,
S. M. Girvin and J. G. E. Harris, \textit{Strong dispersive coupling of a
high-finesse cavity to a micromechanical membrane}, Nature \textbf{452}, 72
(2008).

\bibitem{Chang} D. E. Chang, C. A. Regal, S. B.Pappb, D. J. Wilsonb, J. Ye,
O. Painter, H. J. Kimbleb and P. Zoller, \textit{Cavityopto-mechanics using
an optically levitated nanosphere}, PNAS \textbf{107}, 1005 (2010).

\bibitem{Sankey} J. C. Sankey, C. Yang, B. M. Zwickl, A. M. Jayich and J. G.
E. Harris, \textit{Strong and tunable nonlinear optomechanical coupling in a
low-loss system}, Nature Physics \textbf{6}, 707 (2010).

\bibitem{Peano} V. Peano and M. Thorwart, \textit{Macroscopic quantum
effects in a strongly driven nanomechanical resonator}, Phys. Rev. B \textbf{%
70}, 235401 (2004).

\bibitem{Purdy} T. P. Purdy, D. W. C. Brooks, T. Botter, N. Brahms, Z.-Y. Ma and
D. M. Stamper-Kurn, \textit{Tunable Cavity Optomechanics with Ultracold Atoms%
}, Phys. Rev. Lett. \textbf{105}, 133602 (2010).

\bibitem{Gupta} S. Gupta, K. L.Moore, K. W. Murch and D. M.Stamper-Kurn,
\textit{Cavity Nonlinear Optics at Low Photon Numbers from Collective Atomic Motion},
Phys. Rev. Lett. \textbf{99}, 213601 (2007).

\bibitem{Metzger} C. Metzger, M. Ludwig, C. Neuenhahn, A. Ortlieb, I.
Favero, K. Karrai, F. Marquardt, \textit{Self-Induced Oscillations in an
Optomechanical System Driven by Bolometric Backaction}, Phys. Rev. Lett.
\textbf{101}, 133903 (2008).

\bibitem{Carmon2} T. Carmon, H. Rokhsari, L. Yang, T. J Kippenberg and K. J.
Vahala, \textit{Temporal behavior of radiation-pressure-induced vibrations
of an optical microcavity phonon mode}, Phys. Rev. Lett. \textbf{94}, 223902
(2005).

\bibitem{Okamoto} H. Okamoto, D. Ito, K. Onomitsu, H. Sanada, H. Gotoh, T.
Sogawa, H. Yamaguchi, \textit{Vibration Amplification, Damping, and
Self-Oscillations in Micromechanical Resonators Induced by Optomechanical
Coupling through Carrier Excitation}, Phys. Rev. Lett. \textbf{106}, 036801
(2011).

\bibitem{Zaitsev} S. Zaitsev, A. K. Pandey, O. Shtempluck, and E. Buks,
\textit{Forced and self-excited oscillations of an optomechanical cavity},
Phys. Rev. E \textbf{84}, 046605 (2011).

\bibitem{Lin} L. Zhang and H. Y. Kong, \textit{Self-sustained oscillation and harmonic generation
in optomechanical systems with quadratic couplings}, Phys. Rev. A \textbf{89}, 023847 (2014).

\bibitem{Kippenberg} P. Del¡¯Haye, A. Schliesser, O. Arcizet, T. Wilken, R. Holzwarth and T. J. Kippenberg,
\textit{Optical frequency comb generation from a monolithic microresonator},
Nature \textbf{450}, 1214 (2007).

\bibitem{Jenkins} A. Jenkins, \textit{Self-oscillation}, Phys. Rep. \textbf{525}, 167 (2013).

\bibitem{Long} M. Gao, F. C. Lei, C. G. Du, and G. L. Long,
\textit{Self-sustained oscillation and dynamical multistability of optomechanical systems
in the extremely-large-amplitude regime},
Phys. Rev. A \textbf{91}, 013833 (2015).

\bibitem{Girvin} F. Marquardt, J. G. E. Harris, and S. M. Girvin, \textit{%
Dynamical Multistability Induced by Radiation Pressure in High-Finesse
Micromechanical Optical Cavities}, Phys. Rev. Lett. \textbf{96}, 103901
(2006).

\bibitem{Carmon} T. Carmon, M. C. Cross, and Kerry J. Vahala, \textit{Chaotic
Quivering of Micron-Scaled On-Chip Resonators Excited by Centrifugal Optical
Pressure}, Phys. Rev. Lett. \textbf{98}, 167203 (2007).

\bibitem{Bakemeier} L. Bakemeier, A. Alvermann, and H. Fehske, \textit{Route to Chaos in Optomechanics},
Phys. Rev. Lett. \textbf{114}, 013601 (2015).

\bibitem{Larson} Jonas Larson and Mats Horsdal
\textit{Photonic Josephson effect, phase transitions, and chaos in optomechanical systems},
Phys. Rev. A \textbf{84}, 021804(R) (2011).

\bibitem{Ying} X. L\"{u}, H. Jing, J. Ma, and Y. Wu,
\textit{PT-Symmetry-Breaking Chaos in Optomechanics},
Phys. Re. Lett. \textbf{114}, 253601 (2015).

\bibitem{Karabalin} R. B. Karabalin, X. L. Feng, and M. L. Roukes,
\textit{Parametric Nanomechanical Amplification at Very High Frequency},
Nano Lett. \textbf{9}, 3116 (2009).

\bibitem{Girvin0} A. Nunnenkamp, K. B{\o}rkje, J. G. E. Harris and S. M.
Girvin, \textit{Cooling and squeezing via quardratic optomechanical coupling%
}, Phys. Rev. A \textbf{82}, 021806(R) (2010).

\bibitem{Heidmann} O.Arcizet, P. -F. Cohadon, T. Briant, M. Pinard and A.
Heidmann, \textit{Radiation-pressure cooling and optomechanical instability
of a micromirror}, Nature \textbf{444} 71 (2006).

\bibitem{Aldridge} J. S. Aldridge and A. N. Cleland,
\textit{Noise-Enabled Precision Measurements of a Duffing Nanomechanical Resonator},
Phys. Rev. Lett. \textbf{94}, 156403 (2005).

\bibitem{Katz} I. Katz, A. Retzker, R. Straub, and R. Lifshitz,
\textit{Signatures for a Classical to Quantum Transition of a Driven Nonlinear Nanomechanical Resonator},
Phys. Rev. Lett. \textbf{99}, 040404 (2007).

\bibitem{Venstra} Warner J. Venstra1
, Hidde J. R. Westra, Herre S.J. van der Zant,
\textit{Stochastic switching of cantilever motion},
Nat. Commu. \textbf{4}, 2624 (2013).

\bibitem{Favero} I. Favero and K. Karrai, \textit{Optomechanics of
deformable optical cavities}, Nature Photonics \textbf{3}, 201 (2009).

\bibitem{Walls} D.F. Walls, G.J. Milburn., Chapter7, \textit{Quantum Optics}
2nd edn (Berlin:Springer) 2008.
%

\bibitem{Rugar} D. Rugar, P. Gr\"{u}tter, \textit{Mechanical parametric
amplification and thermomechanical noise squeezing}, Phys. Rev. Lett.
\textbf{67}, 699 (1991).

\bibitem{Turner} K. L. Turner, S. A. Miller, P. G. Hartwell, N. C.
MacDonald, S. H. Strogatz, S. G. Adams, \textit{Five parametric resonances
in a micromechanical system}, Nature \textbf{396}, 149 (1998).

\bibitem{Junho} Junho Suh, M. D. LaHaye, P. M. Echternach, K. C. Schwab and
M. L. Roukes, \textit{Parametric Amplification and Back-Action Noise
Squeezing by a Qubit-Coupled Nanoresonator} Nano Lett. \textbf{10}, 3990
(2010).

\bibitem{Ludwig} M. Ludwig, B. Kubala and F. Marquardt, \textit{The
optomechanical instability in the quantum regime}, New J. Phys. \textbf{10},
095013 (2008).

\bibitem{Kozinsky} I. Kozinsky, H. W. Ch.Postma, O. Kogan, A. Husain, and
M.L.Roukes, \textit{Basins of Attraction of a Nonlinear Nanomechanical
Resonator}, Phys. Rev. Lett. \textbf{99}, 207201 (2007).

\bibitem{Dorsel} A. Dorsel, J. D. McCullen, P. Meystre, E. Vignes, and H.
Walther, \textit{Optical Bistability and Mirror Confinement Induced by
Radiation Pressure}, Phys. Rev. Lett. \textbf{51}, 1550 (1983).

\bibitem{Lin2} L. Zhang and Z. Song, \textit{Modification on static responses
of a nano-oscillator by quadratic opto-mechanical couplings},
Sci. China Phys. Mech. Astron. \textbf{57}, 880 (2014).

\bibitem{Merkin} D. R. Merkin, Introduction to the theory of stability,
Springer-Verlag New York, Inc. 1997, Chapter4 P111.

\bibitem{Edmund} E. X. DeJesus, C. Kaufman, Phys Rev. A \textbf{35}, 5288
(1987).

\bibitem{Lamb} J. S. W. Lamb, and J. A. G. Roberts,
\textit{Time-reversal symmetry in dynamical systems: A survey},
Physica D \textbf{112}, 1 (1998).


\bibitem{Zadeh} M. Hossein-Zadeh and K. J. Vahala,
\textit{Observation of optical spring effect in a microtoroidal optomechanical resonator},
Opt. Lett. \textbf{32}, 1611 (2007).

\bibitem{Sheard} B. S. Sheard, M. B. Gray, C. M. Mow-Lowry, D. E. McClelland and S. E. Whitcomb,
\textit{Observation and characterization of an optical spring},
Phys. Rev. A \textbf{69}, 051801(R) (2004).

\bibitem{Pikovsky} A. Pikovsky, M. Rosenblum and J. Kurths, \textit{Synchronization:
a universal concept in nonlinear sciences}, Cambridge
University Press, Cambridge UK (2001).

\bibitem{Seok} H. Seok, E. M. Wright, and P. Meystre,
\textit{Dynamic stabilization of an optomechanical oscillator},
Phys. Rev. A 90, 043840 (2014).

\bibitem{Milburn} G. J. Milburn, D. F. Walls,
\textit{Quantum nondemolition measurements via quadratic coupling},
 Phys Rev A \textbf{28}, 2065 (1983).

\bibitem{Vanner} M. R.Vanner,
\textit{Selective Linear or Quadratic Optomechanical Coupling via Measurement},
Phys. Re. X \textbf{1}, 021011 (2011).


\end{thebibliography}
\end{document}